\documentclass[twocolumn, twocolappendix]{aastex631}

\usepackage{comment}
\usepackage{hyperref}

\definecolor{indigo}{rgb}{0.0, 0.25, 0.42}
\definecolor{forestgreen}{rgb}{0.13, 0.55, 0.13}

\shorttitle{Modest dust settling in the IRAS04302+2247 Class I protoplanetary disk}
\shortauthors{Villenave et al.}
\graphicspath{{./}{figures/}}

\begin{document}

\title{Modest dust settling in the IRAS04302+2247 Class I protoplanetary disk}

\email{marion.f.villenave@jpl.nasa.gov}
\author[0000-0002-8962-448X]{M.~Villenave}
\affiliation{Jet Propulsion Laboratory, California Institute of Technology, 4800 Oak Grove Drive, Pasadena, CA 91109, USA}

\author[0000-0003-2733-5372]{L. Podio}
\affiliation{INAF – Osservatorio Astrofisico di Arcetri, Largo E. Fermi 5, 50125 Firenze, Italy}

\author[0000-0002-2805-7338]{G. Duch\^ene}
\affiliation{Astronomy Department, University of California, Berkeley, CA 94720, USA}
\affiliation{Univ. Grenoble Alpes, CNRS, IPAG, F-38000 Grenoble, France}

\author[0000-0002-2805-7338]{K. R. Stapelfeldt}
\affiliation{Jet Propulsion Laboratory, California Institute of Technology, 4800 Oak Grove Drive, Pasadena, CA 91109, USA}

\author[0000-0001-9834-7579]{C. Melis}
\affiliation{Center for Astrophysics and Space Sciences, University of California, San Diego, La Jolla, CA 92093-0424, USA}

\author[0000-0003-2862-5363]{C. Carrasco-Gonzalez}
\affiliation{Instituto de Radioastronomía y Astrofísica (IRyA-UNAM), Morelia, Mexico}

\author[0000-0002-5714-799X]{V. J. M. Le Gouellec}
\affiliation{SOFIA Science Center, USRA, NASA Ames Research Center, Moffett Field, CA 94045, USA}

\author[0000-0002-1637-7393]{F. M\'enard}
\affiliation{Univ. Grenoble Alpes, CNRS, IPAG, F-38000 Grenoble, France}

\author[0000-0001-5659-0140]{M. de Simone}
\affiliation{European Southern Observatory, Karl-Schwarzschild-Str. 2,85748 Garching, Germany}

\author{C. Chandler}
\affiliation{ National Radio Astronomy Observatory, PO Box O, Socorro, NM 87801,
USA}

\author[0000-0002-4266-0643]{A. Garufi}
\affiliation{INAF – Osservatorio Astrofisico di Arcetri, Largo E. Fermi 5, 50125 Firenze, Italy}

\author[0000-0001-5907-5179]{C. Pinte}
\affiliation{School of Physics and Astronomy, Monash University, Clayton Vic 3800, Australia}
\affiliation{Univ. Grenoble Alpes, CNRS, IPAG, F-38000 Grenoble, France}

\author[0000-0001-9249-7082]{E. Bianchi}
\affiliation{Excellence Cluster ORIGINS, Boltzmannstraße 2, D-85748 Garching bei München, Germany}

\author[0000-0003-1514-3074]{C. Codella}
\affiliation{INAF – Osservatorio Astrofisico di Arcetri, Largo E. Fermi 5, 50125 Firenze, Italy}
\affiliation{Univ. Grenoble Alpes, CNRS, IPAG, F-38000 Grenoble, France}

\begin{abstract}
We present new VLA observations, between 6.8mm and 66mm, of the edge-on Class~I disk IRAS04302+2247. Observations at 6.8mm and 9.2mm lead to the detection of thermal emission from the disk, while shallow observations at the other wavelengths are used to correct for emission from other processes. 
The disk radial brightness profile transitions from broadly extended in previous ALMA 0.9mm and 2.1mm observations to much more centrally brightened at 6.8mm and 9.2mm, which can be explained by optical depth effects.  
The radiative transfer modeling of the 0.9mm, 2.1mm, and 9.2mm data suggests that the grains are smaller than 1cm in the outer regions of the disk and allows us to obtain the first lower limit for the  scale height of grains emitting at millimeter wavelengths in a protoplanetary disk. We find that the millimeter dust scale height is between 1au and 6au at a radius 100au from the central star, while the gas scale height is estimated to be about 7au, indicating a modest level of settling. The estimated dust height is intermediate between less evolved Class 0 sources, that are found to be vertically  thick, and more evolved Class II sources, which show a significant level of settling. This suggests that we are witnessing an intermediate stage of dust settling. 
\end{abstract}


\section{Introduction} \label{sec:intro}

As the birth place of planets, protoplanetary disks are key to understand the diversity of the observed exoplanet population. Within disks, submicron sized particles grow to large pebbles sizes, that eventually aggregate to form planetesimals or planetary cores~\citep{Drazkowska_2022}. This process can be accelerated in high dust density regions, such as radial substructures or vertically thin dust layers. While radial substructures have been detected in numerous disks~\citep{Andrews_2018}, fewer measurements of the vertical extent of millimeter dust in protoplanetary disks have been performed~\citep[e.g.,][]{Pinte_2016, Doi_Kataoka_2021, Villenave_2022}.

The vertical thickness of dust in a disk is set by the efficiency of vertical settling. This mechanism allows dust particles to concentrate into the midplane, with concentrations that depend on their interaction with the gas~\citep{Weidenschilling_1977}. Larger particles are expected to be more decoupled from the gas and more affected by vertical settling. These large dust grains (e.g., mm sized)  will thus be more concentrated into the midplane than smaller particles (e.g., $\mu$m sized) that remain well mixed with the gas and up to high altitudes~\citep{Barriere-Fouchet_2005}. 
In addition, vertical settling efficiency depends on the turbulence level of the disk, and on its evolutionary stage. 

Star formation is divided into several classes, where Class~0 corresponds to embedded protostars, Class I objects present both a disk and a prominent envelope around the central star, and Class II only have a disk left~\citep[e.g.,][]{Andre_2000}. Direct measurements of millimeter dust scale height in several Class I/II and Class~II disks found that the outer regions are very settled, with a typical scale height of less than 1au at a radius of 100au \citep{Pinte_2016, Villenave_2020, Doi_Kataoka_2021, Villenave_2022, Liu_2022}. In contrast, observations of younger systems such as HH212, VLA 1623 West, and L1527, in the Class~0 and 0/I stage, revealed much thicker disks, possibly not affected by vertical settling~\citep[][]{Lin_2021, Lee_2022, Michel_2022, Ohashi_2022, Sheehan_2022, Sakai_2017}. Further constraints on the evolution of the vertical extent of protoplanetary disks of different evolutionary stages, specifically in the Class I, are important to determine the efficiency of this mechanism with age.

In this work, we focus on the IRAS04302+2247 protoplanetary disk system (hereafter IRAS04302) located in the L1536 cloud in the Taurus star-forming region~\citep[$d=161\pm3$\,pc,][]{Galli_2019}.
This disk, classified as Class I by \citet{Kenyon_1995}, has been the subject of a number of studies at various wavelengths providing useful insights on its structure. The scattered light images~\citep{Lucas_Roche_1997, Padgett_1999} show a clear dark lane, indicating the presence of a disk seen almost edge on \citep[$90\pm3$\,\degr,][]{Wolf_2003}, and a bipolar nebula, dominated by a prominent envelope structure. The source was also observed at millimeter wavelengths, both in continuum and in different molecular lines~\citep{Wolf_2008, Podio_2020, vant_Hoff_2020, Villenave_2020}.  The Atacama Large Millimeter Array (ALMA) observations at 2.1mm obtained by \citet{Villenave_2020} resolved the minor axis size of the disk (beam size $0\farcs09\times0\farcs04$) and show that the disk is flared, i.e., with a minor axis width increasing with distance from the star. Here, we present new Karl G. Jansky Very Large Array (VLA) observations at wavelengths between 6.8mm and 66mm. These observations are expected to probe grains that are up to one order of magnitude larger than those probed by previous ALMA observations, which are predicted to be more affected by vertical settling. 
We present the observations in Sect.~\ref{sec:obs} and observational results in Sect.~\ref{sec:results}. Then, in Sect.~\ref{sec:model}, we compare 0.9mm, 2.1mm, and 9.2mm observations to a grid of radiative transfer models with the aim to obtain constraints on the vertical and radial distribution of the largest grains in the disk. The results are discussed in Sect.~\ref{sec:discussion}, and we conclude this work in Sect.~\ref{sec:conclusion}.

\section{Observations and data reduction}
\label{sec:obs}

In this section, we present the new VLA  observations of IRAS04302 at 6.8mm, 9.2mm, 14mm, 40mm, and 66mm. In addition, we also use previously published photometric data points from \citet{Grafe_2013}, and ALMA 0.9mm and 2.1mm observations from \citet{Villenave_2020}. We refer the reader to \citet{Villenave_2020} for details on the calibration of the ALMA observations used in this work.

\subsection{VLA Ka data}
\label{sec:VLA_Ka}
IRAS04302 was observed in the Ka band (9.2mm, Project: 21B-183, PI: Podio) in September and October 2021 with the VLA B configuration for a total of 8.5h on source. 
In Table~\ref{tab:vla-obs}, we report the observing dates, frequency range, as well as the flux and phase calibrators used for these observations.  

\begin{table*}[]
    \centering
    \caption{VLA observation log table.}
    \begin{tabular}{ccccccclccc}
    \hline\hline
     B & $\lambda$ & $\nu$  & Beam size &  Beam PA & RMS & TToS & Project & Observation  & Flux & Phase\\
    &(mm) & (GHz)& ($\arcsec$) & ($^\circ$) & ($\mu$Jy/beam) & (min) &&date & calibrator & calibrator\\
    \hline
    Q & 6.8 & $40-48$ & $0.19\times0.15$ & $-6$ & 15.8 & 45& 21B-100 & 2021/12/18 + 21 &3C48 &J0431+2037\\ 
    Ka & 9.2 & $28-29$ & $0.19\times0.18$& 87& 4.3 &510 &21B-183 & 2021/09/30 & 3C147 & J0426+2327\\ 
     & &+$36-37$ & & &&&  & 2021/10/04 & 3C286 & J0426+2327\\
     & & & & &&& & 2021/10/21 & 3C147 & J0426+2327\\
     & & & & &&&  & 2021/10/22 & 3C286 & J0426+2327\\
    K & 14 & $18-23$& $0.37\times0.29$ &  $-14$& 8.3 & 10& 21B-100 & 2021/12/18 + 21 &3C48 &J0431+2037\\
    C1 & 40 & $7-8$& $1.07\times0.88$ & $0$ & 9.4  & 8& 21B-100 & 2021/12/18 + 21 & 3C48&J0431+2037\\
    C2 & 66 & $4-5$& $1.70\times1.42$ & $-2$ & 10.6  & 8& 21B-100 & 2021/12/18 + 21 &3C48 &J0431+2037\\
    \hline
    \end{tabular}
    \tablecomments{B: Observing band; $\lambda$: Wavelength corresponding to the average frequency representative of the image;  $\nu$: frequency range of the observations; TToS: Total time on source  }
    \label{tab:vla-obs}
\end{table*}
We calibrated the raw data using the CASA VLA data reduction pipeline version 6.2.1.7~\citep{CASA_2022}. Before producing the final images, we identified a mismatch in flux density of a factor 1.07 between observations taken on September 30/October 21 and those taken on October 04/October 21. The observations made on September 30 and October 21 used the flux calibrator 3C147, which is known to be variable (see \href{https://science.nrao.edu/facilities/vla/docs/manuals/oss/performance/fdscale}{VLA manual}), contrary to 3C286 which was used for the other two observations in Ka band. We thus adjusted the fluxes of the observations from September~30 and October 21 to match those of October 04 and 22 using the \texttt{gencal} and \texttt{applycal} CASA tasks. To maximize the dynamic range of the image, we then performed phase self-calibration on the observations. Finally, we produced the final continuum image using the \texttt{tclean} task, with a briggs weighting (robust 0.5). The resulting continuum beam size and rms are reported in Table~\ref{tab:vla-obs}.

\begin{figure*}
    \centering
    \includegraphics[width = 1\textwidth]{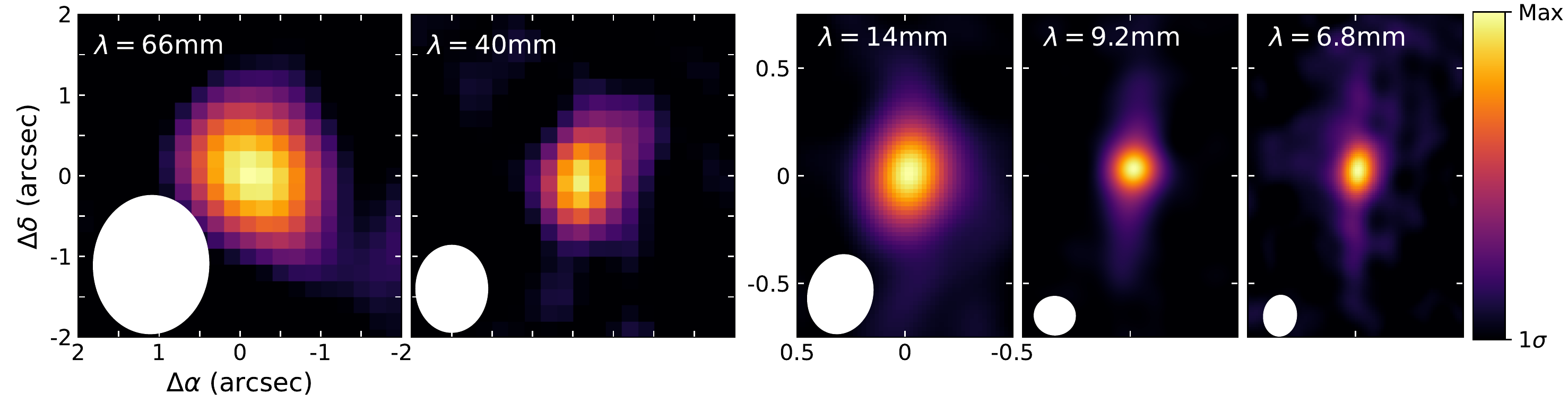}
    \caption{VLA observations of IRAS04302. The scale bar represents the flux level; its values range between the 1$\sigma$ level (RMS; Table~\ref{tab:vla-obs}) and the peak flux of each map. This corresponds to a brightness ratio (Max/1$\sigma$) of 7, 15, 34, 71, and 22, respectively for the 66mm, 40mm, 14mm, 9.2mm, and 6.8mm data.  We show the beam size (ellipse) in the bottom left corner of each panel.}
    \label{fig:VLA_obs}
\end{figure*}

The 9.2mm image  resulting directly from the data is presented in Fig.~\ref{fig:VLA_obs}. It is likely that the central beam ($\sim0\farcs2$) includes contribution from  processes other than dust thermal emission, such as free–free emission from ionized jets or disk winds and gyrosynchrotron emission from coronal processes~\citep[see][and references therein]{Melis_2011}. Throughout the text we also refer to these processes under the terms ``non-dust emission" or ``wind/coronal emission". 
To subtract such non-dust emission, we use other available VLA data (see Sect.~\ref{sec:data_QKC}) and follow the method previously implemented by \citet{Carrasco-Gonzalez_2019}. We describe the methodology in detail in Appendix~\ref{app:free-free}. 
Ultimately in this paper when modeling the disk (Sect.~\ref{sec:model}) we consider only the regions outside of 1 beam size to limit the effect of potential variability of processes other than continuum dust emission. 

\subsection{VLA Q, K, and C data}
\label{sec:data_QKC}

IRAS04302 was also observed in Q, K, and C bands (Project: 21B-100, PI: Villenave), corresponding to wavelengths between 6.8mm and 66mm,  in December 2021 with the VLA in the B configuration. Table~\ref{tab:vla-obs} presents the frequency range, time on source, observing dates, and calibrators used for these observations. We note that the C band observations were performed with two widely separated basebands of 1 GHz bandwidth each, which we separated into C1 (40mm) and C2 (66mm) to increase the wavelength range probed by these observations. We calibrated the raw data using the CASA pipeline version 6.2.1.7.  
To increase the signal to noise of the emission, we produced the final images using the \texttt{tclean} task with a natural weighting. We report the beam sizes and rms in Table~\ref{tab:vla-obs}, and show the final images in Fig.~\ref{fig:VLA_obs}. Similarly to the observations at 9.2mm, the central regions of the 6.8mm observations are likely affected by processes other than dust thermal emission, which we correct following the methodology described in Appendix~\ref{app:free-free}.

\section{Results}
\label{sec:results}
\subsection{Continuum emission, fluxes, and sizes}

\begin{figure}
    \centering   
    \includegraphics[width = 0.5\textwidth, trim={0.cm 0cm 0cm 0.cm}, clip]{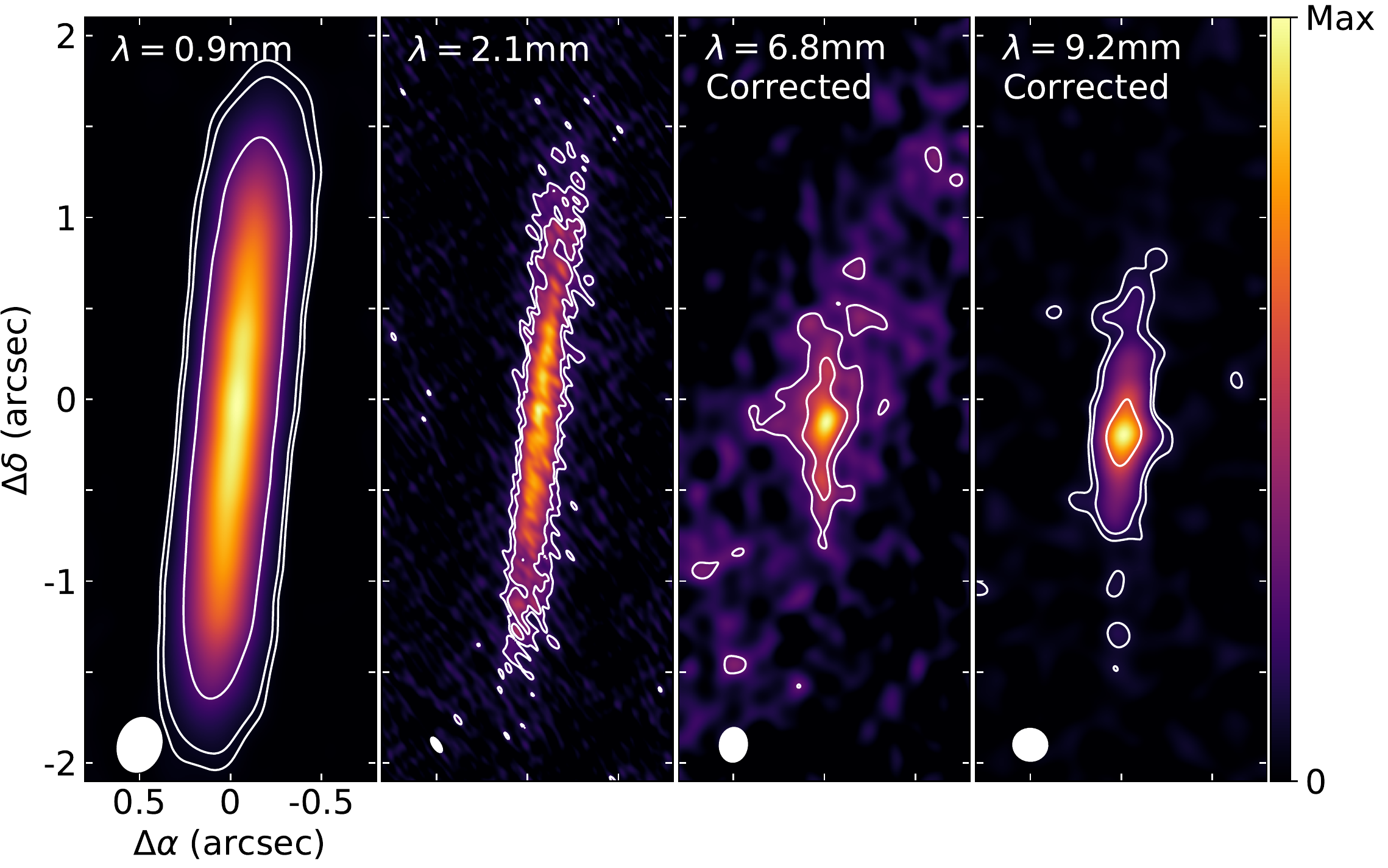}
    
    \caption{Side by side comparison of the ALMA 0.9mm, 2.1mm and VLA 6.8mm and 9.2mm observations, after correction for emission from processes other than dust continuum emission (Appendix~\ref{app:free-free}). The contours indicate emission at the 3, 5, and 20$\sigma$ levels of each map. The signal to noise of each map (Max/1$\sigma$) is  83, 17, 11, and 34, respectively for the  0.9mm, 2.1mm and VLA 6.8mm and 9.2mm data. Because of the correction for non-dust emission, the signal to noise of the 6.8mm and 9.2mm data is decreased compared to that reported in Fig.~\ref{fig:VLA_obs}.
    \label{fig:ALMA_VLA}}
\end{figure}

In Fig.~\ref{fig:VLA_obs}, we present the new 66mm, 40mm, 14mm, 9.2mm, and 6.8mm observations without any correction for coronal/wind emission. We find that the 66mm, 40mm, and 14mm observations are unresolved. The spectral index obtained for these three wavelengths (see Sect.~\ref{sec:spectral_indices}) is inconsistent with dust emission, so we conclude that they are dominated by coronal/wind emission. On the other hand, the 9.2mm and 6.8mm observations are strongly centrally peaked, with detection of extended structure along the disk major axis. The extended component corresponds to thermal emission from the disk, while the central beam is  dominated by emission from other processes. 
In Fig.~\ref{fig:ALMA_VLA}, we show the ALMA 0.9mm and 2.1mm, along with the VLA 6.8mm and 9.2mm observations after correction for processes other than dust continuum emission (see Appendix~\ref{app:free-free}). Even after correction, we find that the emission at 6.8mm and 9.2mm is centrally peaked (see also Fig.~\ref{fig:spectral_index}). This could suggest a higher dust concentration or temperature within the inner region of the disk, although caution is needed due to the potential variability of free-free and gyrosynchrotron emission.\\

\begin{table}[]
    \begin{centering}
    \caption{Fluxes and major axis sizes of the VLA and ALMA observations. \label{tab:flux}}
    \begin{tabular}{ccccc}
    \hline\hline
    $\lambda$ (mm) & Flux  ($\mu$Jy) & $R_{68\%}$ ('') & $R_{95\%}$ ('')\\
    \hline
    0.9  & $267500 \pm 26752$& 0.90 &  1.52\\
    2.1 &$44500 \pm 4450^{a}$ & 0.72 & 1.51 \\
    6.8   & $1850 \pm 186^{b, *}$ & 0.52 & 1.37 \\
    9.2  & $820 \pm 82^{b}$ & 0.39 & 1.01 \\
    14  &$294 \pm 17$ & -& -\\
    40  &$139 \pm 11$ & -& -\\
    66  &$62 \pm 11$ & -& -\\
    \hline
    \end{tabular}
    \end{centering}
    \tablecomments{The major axis sizes at 6.8mm and 9.2mm were measured on the images corrected for emission from processes other than dust continuum emission. $^{(a)}$ The 2.1mm flux reported here is $\sim 20\%$ higher than the flux reported in \citet{Villenave_2020}, possibly due to differences in aperture sizes. $^{(b)}$ After subtracting for non-dust emission, the total flux at 6.8mm and 9.2mm are $1690 \pm 170 \mu$Jy and $654 \pm 65 \mu$Jy, respectively. $^{(*)}$ The flux density of the low signal to noise 6.8mm data strongly depends on the aperture size, likely due to the presence of large scale background structure. It can be decreased by a factor of nearly 2 if the aperture size is smaller. }
\end{table}

We present the flux density of the source at the different wavelengths in Table~\ref{tab:flux}. For the observations where the source is unresolved, namely the 14mm, 40mm, and 66mm observations, we fitted the visibilities by a point source using the \texttt{uvmodelfit} CASA task. On the other hand, we estimate the flux density of the extended 6.8mm and 9.2mm observations in the images using an aperture of $0\farcs8\times4''$ ($129\times644$au). We also re-estimate the 2.1mm and 0.9mm fluxes from the ALMA observations of \citet{Villenave_2020} using the same aperture, which encompasses all the emission at the four bands. 
For all wavelengths, we evaluate the uncertainty by taking the quadratic sum of the rms (Table~\ref{tab:vla-obs}) and systematic uncertainty of the observatory (10\% for $\lambda \leq 13 \mu$m and 5\% otherwise, see \href{https://science.nrao.edu/facilities/vla/docs/manuals/oss/performance/fdscale}{VLA manual}). \\

We estimate the disk size using the cumulative flux technique along the major axis. Previous literature works~\citep[e.g.,][]{Ansdell_2018, Long_2018}, focusing mostly on low- to mid-inclination systems, generally estimated the cumulative flux using elliptical apertures (related to the disk inclination) of increasing radii. However, IRAS04302 is highly inclined and typically unresolved in the minor axis direction, thus using an elliptical aperture to measure the disk size can bias the results. A better approach is to estimate the radius from the cumulative flux obtained from the major axis cut directly  \citep[see also][who made a similar choice when measuring the gas size of the edge-on disk around HK\,Tau\,B]{Rota_2022}.
We calculate the cumulative flux at increasingly larger radii along the major axis cut, where the major axis cut corresponds to a 1-pixel line along the position angle (PA) of the disk and passing by the continuum peak. We then estimate the disk radius $R_{95\%}$ as the radius containing 95\% of the total flux, and the effective disk radius as that containing 68\% of the total flux ($R_{68\%}$). We note that for 6.8mm and 9.2mm, we use the maps corrected for coronal/wind emission to estimate the disk size. However we checked that using the non-corrected maps only lead to minimal changes in the apparent size, of order $\lesssim0\farcs1$. To obtain consistent comparison between the new centimeter range and previous millimeter range observations, we also apply this technique to the previously published ALMA 2.1mm and 0.9mm observations \citep{Villenave_2020}. We report the results in Table~\ref{tab:flux}. We find that both the 68\% and the 95\% flux radii decrease with wavelength. This is also visible on the top panel of Fig.~\ref{fig:spectral_index}, which shows the major axis profiles of the 0.9mm, 2.1mm, 6.8mm, and 9.2mm observations. 

\subsection{Spectral indices}
\label{sec:spectral_indices}

The spectral index of millimeter-centimeter emission, defined as $\alpha$ in $F_\nu\propto \nu^\alpha$, can be used to study the dust  and optical depth properties within a disk. From our new VLA fluxes at 66mm, 40mm, and 14mm (see Table~\ref{tab:flux}), we estimate a spectral index of $\alpha_{14-66mm} = 0.97\pm0.24$. Using the integrated fluxes reported in Table~\ref{tab:flux} and from \citet{Grafe_2013}, we also estimate a spectral index between 0.9mm and 9.2mm of $\alpha_{0.9-9.2mm}= 2.54\pm 0.06$. At longer wavelengths, the spectral index is dominated by coronal/wind emission, while dust dominates at shorter wavelengths. 

\begin{figure}
    \centering
    \includegraphics[width = 0.45\textwidth]{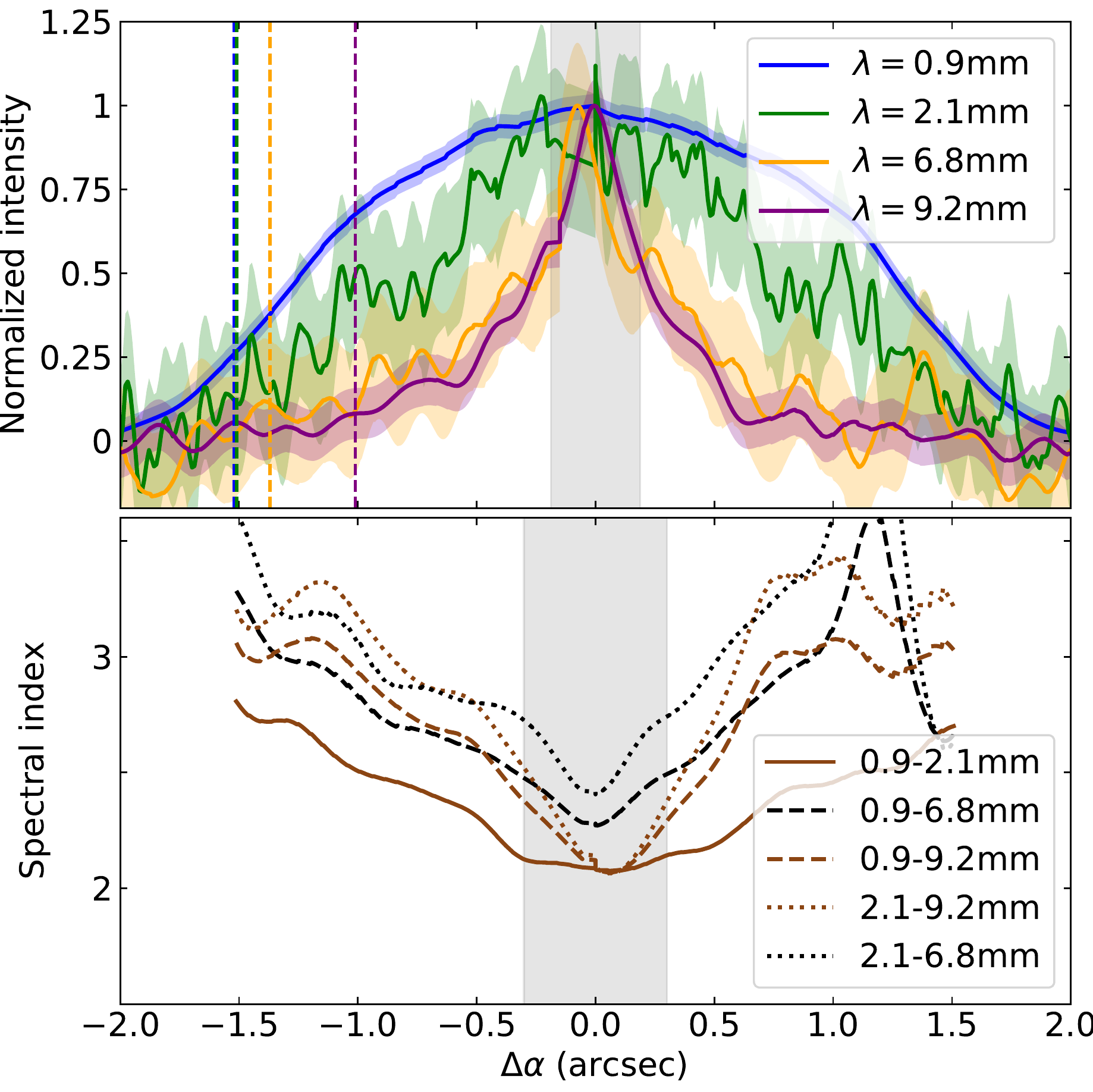}
    \caption{Top panel: Major axis profiles of the 0.9mm, 2.1mm, 6.8mm, and 9.2mm observations (Fig.~\ref{fig:ALMA_VLA}), normalized to their peak. The vertical lines illustrate the location of $R_{95\%}$ (Table~\ref{tab:flux}). Bottom panel: Major axis profiles of the spectral index maps (Fig.~\ref{fig:spectral_index_maps}). The grey region in both panels indicates the region possibly still impacted by coronal/wind emission. It is larger in the bottom plot because the 6.8mm and 9.2mm data were generated at the larger resolution of the 0.9mm image.}
    \label{fig:spectral_index}
\end{figure}

\begin{figure}
    \centering
    \includegraphics[width = 0.5\textwidth]{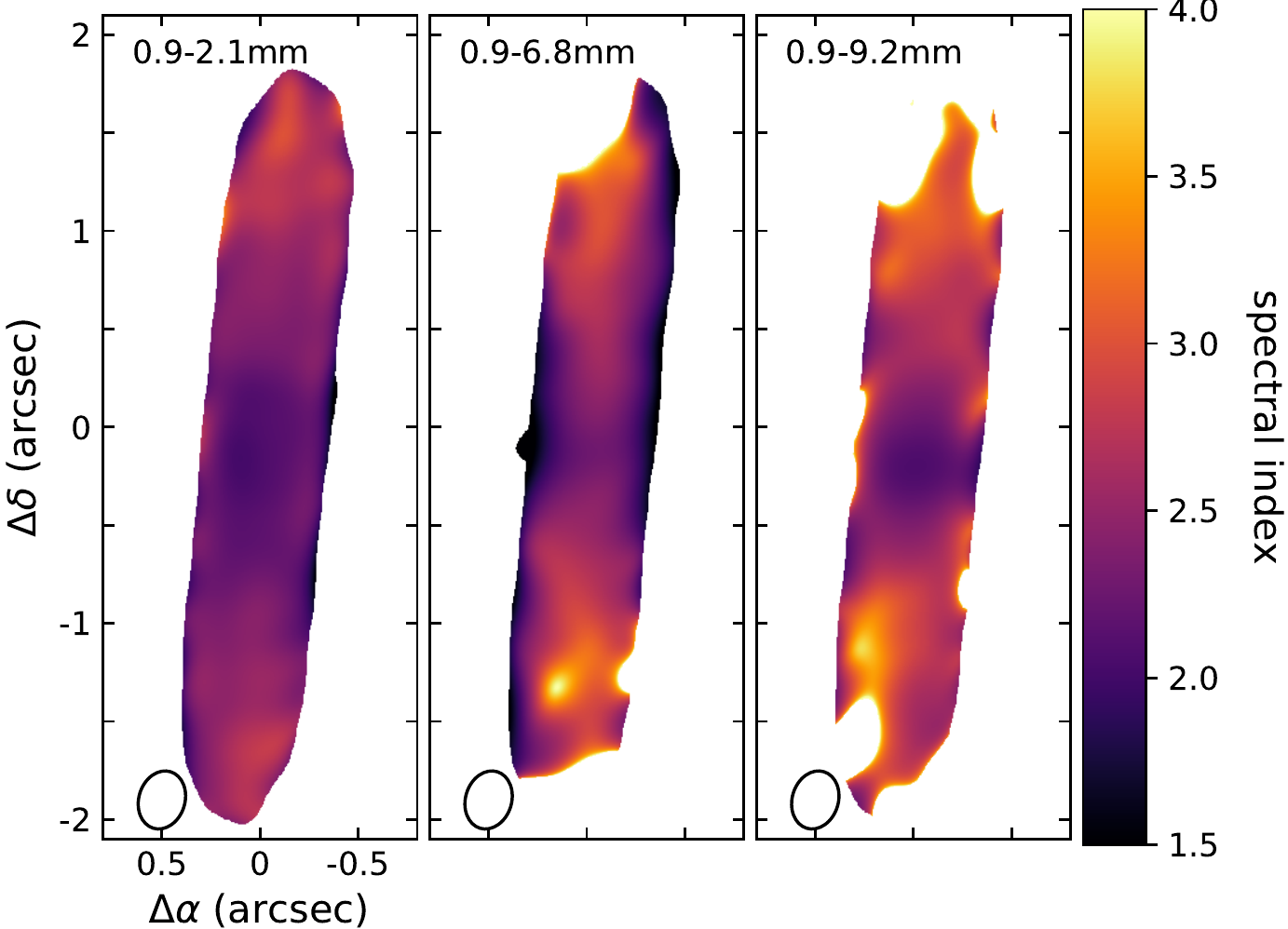}
    \caption{Centimeter and millimeter spectral index maps.}
    \label{fig:spectral_index_maps}
\end{figure}

We also compute the spectral index maps of the centimeter-millimeter emission of IRAS04302. To do so, we first generated 6.8mm and 9.2mm images at the same resolution as the 'restored' resolution used by \citet[][$0\farcs30 \times 0\farcs24$, see their Table A.1]{Villenave_2020} with the \texttt{imsmooth} CASA task. We then generated spectral index maps using the \texttt{immath} CASA task on the images at the same resolution. We present the resulting maps in Fig.~\ref{fig:spectral_index_maps}, and the major axis cuts in Fig.~\ref{fig:spectral_index}.

We find that spectral indices increase at larger radii, and also at longer wavelengths. This behaviour is consistent with the expected decrease of optical depth to outer radii. At the center of the disk, emission is presumably very optically thick, and we obtain a spectral index of 2 at nearly all wavelengths. This suggests that the subtraction of the emission from processes other than dust continuum emission (Appendix~\ref{app:free-free}) was reasonable, as wind/coronal emission is generally associated with lower spectral index~\citep[e.g.,][]{Rodmann_2006}. 
When using the longer wavelengths (9.2mm), we see a steeper decrease of the spectral index within 0.5\arcsec\ from the center of the disk. This would also be expected in the case of a concentration of larger particles at the center of the disk. However, given the high optical depths expected at these radii, this cannot be clearly concluded from this simple analysis.

\section{Radiative Transfer Modeling}
\label{sec:model}
\subsection{Methodology}
\subsubsection{Model description}
\label{sec:description}

To characterize the distribution of large dust in IRAS04302, we model the new 9.2mm VLA and previous 0.9mm and 2.1mm ALMA observations of the system with the radiative transfer code \texttt{mcfost}~\citep{Pinte_2006, Pinte_2009}. \texttt{Mcfost} solves the temperature structure using Monte-Carlo methods based on the disk structure and dust properties, properly accounting for scattering effects. We do not consider the 6.8mm data in this analysis because they have a lower signal to noise and same angular resolution as the 9.2mm data (see Table~\ref{tab:vla-obs}).

We assume that the disk is axi-symmetric, smooth, and that the surface density follows a power-law distribution: $\Sigma(r)\propto r^{-p}$, for $R_{in}<r<R_{out}$. The vertical extent of the grains is parameterized such that $ H(r) = H_{100au}(r/100au)^\beta$. We assume that the grain size follows a power law distribution such that $n(a)da\propto a^{-3.5}da$.  

In addition, as described in Appendix~\ref{appx:match_thermal_SED}, we implement a simplified parametrization of vertical settling, namely with a disk represented by a layer of small grains located up to high altitudes in the disk (characterized by $H_{sd, 100au}$), and a layer of larger dust more concentrated towards the midplane~\citep[characterized by $H_{ld, 100au}$, see e.g.,][]{Villenave_2022}. The complete model, described in  Appendix~\ref{appx:match_thermal_SED}, also includes an envelope to reproduce the thermal part of the spectral energy distribution.

\subsubsection{Fixed and varied parameters}
\label{sec:param}

We use a stellar effective temperature of $T_\mathrm{eff}=4500$\,K, stellar radius of $3.7 R_\odot$~\citep{Grafe_2013}. We fix the inner disk radius to $R_{in} = 0.1$\,au, and the outer radius to $R_{out}=300$\,au based on the  apparent size of the 0.9mm observations. For the layer of big grains, we fix the minimum grain size to $a_{min} = 10\,\mu$m. Following \citet{Grafe_2013}, we also assume that dust grains are composed of a mixture of 62.5\% of astronomical silicates and 37.5\% of graphite. 

We then construct a grid of models by varying several parameters: the dust scale height $H_{ld, 100au}$, inclination~$i$, maximum grain size $a_{max}$, surface density exponent $p$, flaring exponent $\beta$, and dust mass $M_{dust}$. We report the range of explored parameters in Table~\ref{tab:grid}. We note that we consider only inclinations  between 88$^\circ$ and 90$^\circ$. These values were chosen based on previous modeling of scattered light observations, finding an inclination of $90\pm3^\circ$~\citep[][]{Wolf_2003, Lucas_Roche_1997}, and based on the shape of the 2.1mm image, which implies an inclination $>84^\circ$~\citep{Villenave_2020}. Our initial modeling trying models between 84$^\circ$ and 88$^\circ$ produced only very bad fits and thus we limited our grid to a minimum inclination of 88$^\circ$. 
Our grid consists of $5\times3\times3\times3\times4\times13 =7020$ models. For each set of parameters, we compute the 0.89mm, 2.1mm, and 9.2mm images, and convolve the resulting images by the ALMA or VLA 2D Gaussian beam. 

\begin{table}[]
    \centering
     \caption{Overview of the parameter ranges of the grid.}
    \begin{tabular}{lcc}
    \hline\hline
    $H_{ld, 100au}$ & (au) & [1, 3, 4, 5, 6]\\
    $i$ & ($^\circ$) & [88, 89, 90]\\
    $a_{max}$ & ($\mu$m) & [100, 1000, 10000]\\
    $p$ & ($-$) & [0.5, 1, 1.5]\\
    $\beta$ & ($-$) & [1.06, 1.14,  1.22, 1.3]\\
    $\log_{10}(M_{dust}/M_\odot)$ & ($-$) & [-5.0  , -4.75, -4.5 , -4.25, -4.0 \\ && -3.75, -3.5 , -3.25, -3.0,  -2.75\\ 
    && -2.5 , -2.25, -2.0]  \\
    \hline
    \end{tabular}
    \tablecomments{$H_{ld, 100au}$, $i$, $a_{max}$, $\beta$, $M_{dust}$ are respectively the large dust scale height at 100au, the inclination, the maximum grain size, surface density exponent, flaring parameter, and the dust mass in the large grain layer.}
    \label{tab:grid}
\end{table}

\subsubsection{Fitting procedure}
\label{sec:fitting}

The goal of this modeling is to reproduce the shape of the millimeter and centimeter emission in order to obtain constraints on the vertical extent of the larger dust particles. 
Thus, we first build our models by including only a disk region with large grains ($a_{min} =10 \mu$m, and varied $a_{max}$, see Table~\ref{tab:grid}). While adding a layer of smaller grains and envelope allows for a more realistic disk temperature structure, calculating such models is significantly more computationally expensive than when only including the layer of big grains. In Appendix~\ref{appx:match_thermal_SED}, we checked that adding such layers does not significantly affect the shape of millimeter/centimeter emission, while allowing a good match in the thermal part of the SED between model and data. Our modeling is thus based on a layer of big grains only.

Our modeling strategy follows 2 key steps, summarized hereafter:
\begin{enumerate}
    \item For each set of parameters ($H_{ld, 100au}$, $i$, $a_{max}$, $p$, $\beta$), we first determine the dust mass $M_{dust}$ that offers the best match with the observations along the major axis only. This allows us to obtain optical depths of best-mass models consistent with the observations. 
    \item We then evaluate the agreement of each best-mass model with the data by considering both the major and minor axis profiles at the three different wavelengths. 
\end{enumerate}

For both steps, we determine the best models by considering observations at 0.89mm, 2.1mm, and 9.2mm. 
For each wavelength, we evaluate the model accuracy in reproducing either the major or minor axis profiles of the data by estimating a  $\chi^2$, defined by:
\begin{equation}
    \chi^2 = \sum \frac{1}{ n} \left(\frac{F_{d} - F_{m}}{\sigma}\right)^2 
    \label{eq:chi2}
\end{equation}
where $F_{d}$ and $F_m$ are the respective fluxes of the data and model along the major or minor axis, normalized so that their peak intensity is equal to 1, $\sigma$ is the normalized rms of the data, and $n$ is the number of pixels along the cut. We note that for the 9.2mm observations, the region within one beam from the central star is not considered, and thus we normalize the data and model major axis profiles to 1 at the distance of one beam from the central star. The accuracy of each model along the major axis is then given by $\bar{\chi}^2_{maj}$, taken as the mean $\chi^2_{maj}$ between 0.89mm, 2.1mm, and 9.2mm.  For each set of parameters ($H_{ld, 100au}$, $i$, $a_{max}$, $p$, $\beta$), step~1 determines the best-mass model (best $M_{dust}$), which corresponds to that with the lowest $\bar{\chi}^2_{maj}$.

Then, for step 2, we evaluate the agreement of each best-mass model with the data by considering both the major and minor axis profiles at the three different wavelengths. In addition to $\bar{\chi}^2_{maj}$, we evaluate $\bar{\chi}^2_{min}$ along the minor axis direction. We generate minor axis profiles by taking the mean of the cuts at all distances along the major axis~\citep[see][]{Villenave_2020}, excluding the central region for the 9.2mm comparisons. For each wavelength, we then estimate $\chi^2_{min}$ using equation~(\ref{eq:chi2}) along the minor axis profiles, and obtain $\bar{\chi}^2_{min}$ as the average between 0.89mm, 2.1mm, and 9.2mm. Finally, the global agreement between our model and the data used in step~2 is obtained by $\phi^2 = \frac{\bar{\chi}^2_{maj} + \bar{\chi}^2_{min}}{2}$.

\subsection{Results}
\subsubsection{Grid best models}
\label{sec:grid_models}
\begin{figure}
    \centering
    \includegraphics[width = 0.48\textwidth]{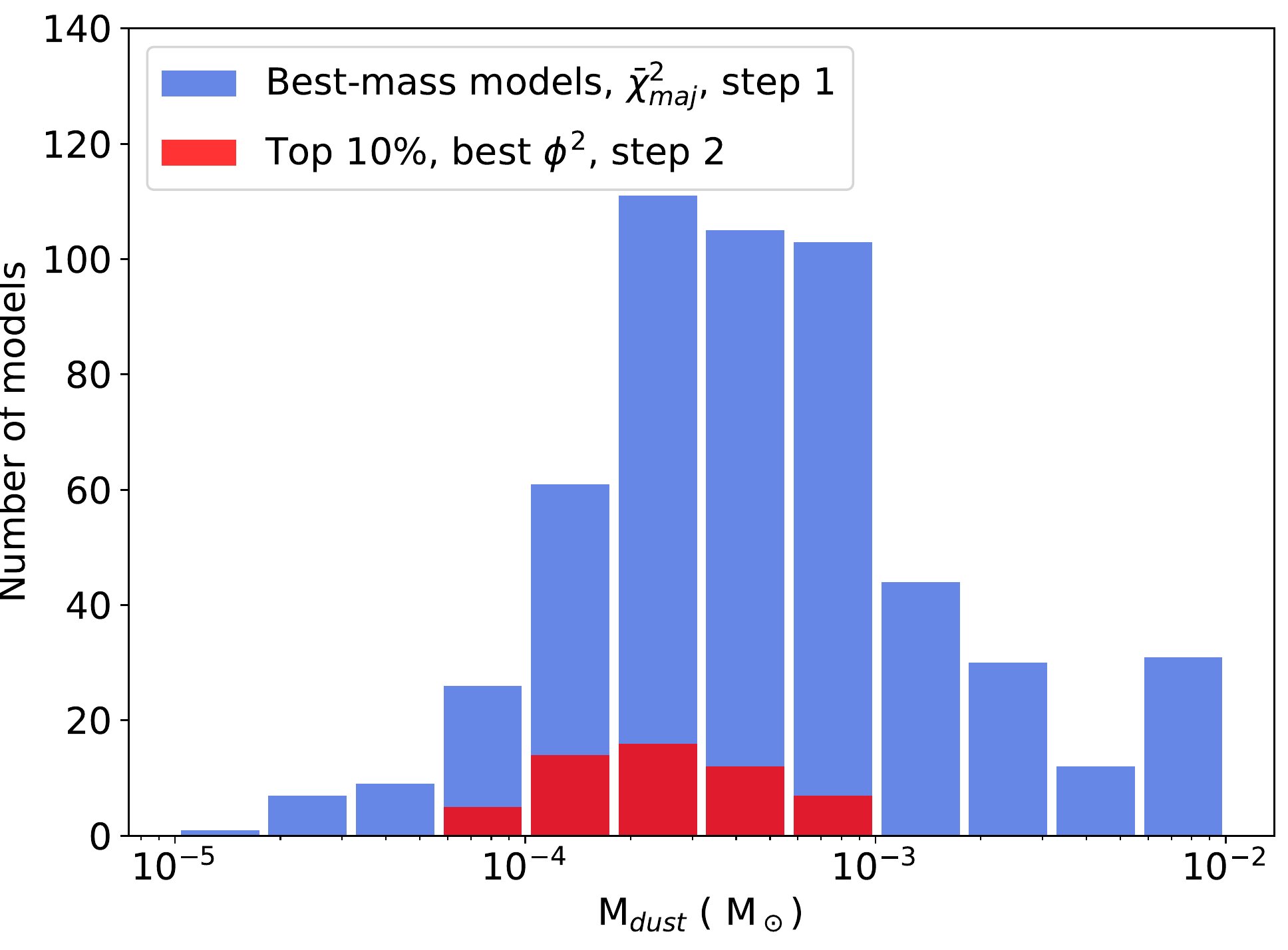}
    \caption{Best dust masses M$_{dust}$ for each combination of ($H_{ld, 100au}$, $i$, $a_{max}$, $p$, $\beta$) in the grid, based on step 1 (using $\bar{\chi}^2_{maj}$, blue bars). For comparison, the red bars show the dust masses of the 10\% best models, based on $\phi^2$. In our initial grid (i.e., before step 1), there are $5\times3\times3\times3\times4=540$ models in each mass bin (Table~\ref{tab:grid}).} 
    \label{fig:masses}
\end{figure}
\begin{figure*}
    \centering
    \includegraphics[width= \textwidth]{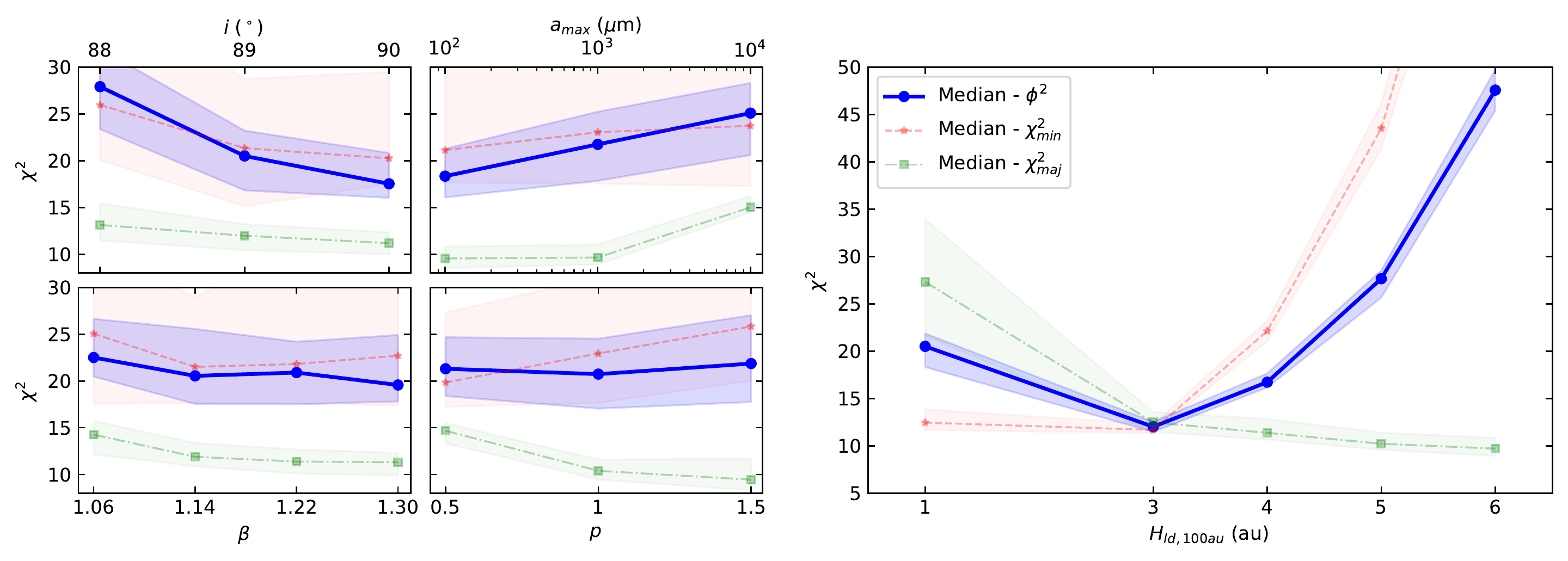}
    \caption{Integrated median $\phi^2$, $\bar{\chi}_{min}^2$, $\bar{\chi}_{maj}^2$ for all best-mass models in our grid. The 40\% and 60\% percentiles are represented by the shaded contours. Models with $\phi^2>25$ are typically bad fits to the data.}
    \label{fig:integrated_grid}
\end{figure*}

We show the distribution of dust masses for each combination of ($H_{ld, 100au}$, $i$, $a_{max}$, $p$, $\beta$), as obtained during step 1, in Fig.~\ref{fig:masses}. The median of the distribution is $M_{dust}= 3.2\times 10^{-4}\ M_\odot$, and the 25\% and 75\% quartiles respectively reach values of $1.8 \times 10^{-4}\ M_\odot$ and $1.8 \times 10^{-3}\ M_\odot$. For comparison, we also indicate the dust masses of the 10\% best models (based on $\phi^2$) in this figure. They fall within the range of best dust masses of the distribution from step 1. 

In Fig.~\ref{fig:masses}, we can also see that 24 models saturate to the high mass limit of our grid, namely at $M_{dust} = 10^{-2} M_\odot$. Those are exclusively models with $H_{ld, 100au} = 1$ au, and $i = 88^\circ$, which are typically more centrally peaked than models at higher inclination or with larger scale height. They require a high dust mass to become sufficiently optically thick to reproduce the observations. In fact, nearly all models with this combination of scale height and inclination are extremely high mass. However, it is very unlikely that the dust mass in IRAS04302 is as high or higher than  $M_d = 10^{-2} M_\odot$, because, assuming a gas to dust ratio of 100, this would imply a star-to-dust mass ratio close to or higher than~1 ($M_\star\sim$ 1.6 $M_\odot$, Lin et al. in preparation). Thus, this first modeling step indicates that IRAS04302 likely does not have the combination of $H_{ld, 100au} = 1$ au and $i = 88^\circ$. \\

After performing both fitting steps, we find that the parameters of our best model are  $H_{ld, 100au} = 3 $au, $i=89^\circ$, $a_{max}=1$mm, $\beta$=1.14, $p=1.5$, and $M_{dust}=3\times 10^{-4} M_\odot$. This model is associated to the following metrics: $\phi^2=8.1$,  $\chi^2_{maj} =5.5$, and $\chi^2_{min}=10.6$. We show the major and minor axis profiles of this best model in Appendix~\ref{appx:match_thermal_SED}, and the maps in Appendix~\ref{appx:additional_models}.\\

In addition, we also compute the mean and median $\phi^2$, $\chi_{min}^2$, $\chi_{maj}^2$ values for each $H_{ld, 100au}$, $i$, $a_{max}$, $p$, $\beta$ value of the grid, and show them in Fig.~\ref{fig:integrated_grid}. 
We find that  $H_{ld, 100au}$ is relatively well constrained, while there are trends for the values of $a_{max}$ and $i$. Specifically, the models indicate that  $1\mathrm{au}<H_{ld, 100au}<6\mathrm{au}$, and suggest that $a_{max}<1$cm and $i>88^\circ$. For the scale height, we note that  $\bar{\chi}_{min}^2$ and $\bar{\chi}_{maj}^2$ favor opposite values, such that the constraint that we obtain is a compromise between the two. 
On the other hand, the profiles for $\beta$ and $p$ appear relatively flat and these parameters are thus not constrained.

\begin{figure*}
    \centering
    \includegraphics[width = 1\textwidth, trim={0.8cm 0cm 0.4cm 0.6cm}, clip]{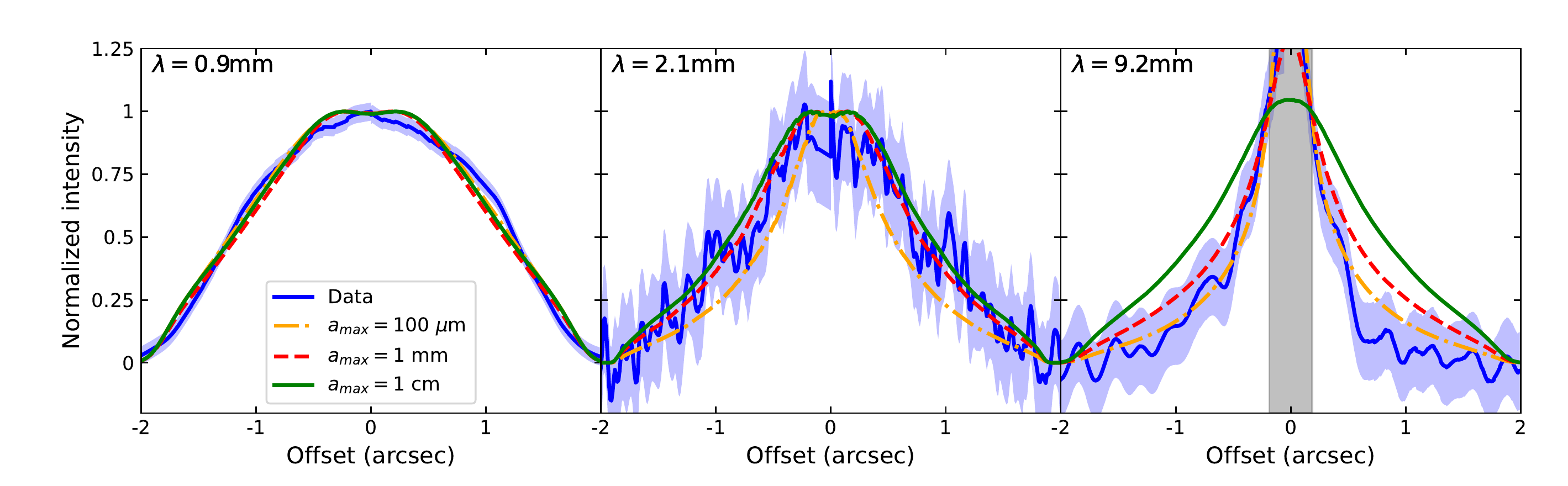}
    \caption{Effect of grain size on the major axis profiles, for models with $H_d=3au$, $i=90^\circ$, $a_{max}$, $p=-1$, and $\beta=1.14$. The shaded contours show 3 times the rms of the observations. The grey region in the 9.2mm panel indicates the zone potentially contaminated by coronal/wind emission, and not considered in this analysis.}
    \label{fig:maj_cut}
\end{figure*}

\subsubsection{Radial extent and maximum grain size}
\label{sec:major_profiles}

As illustrated in Fig.~\ref{fig:integrated_grid}, the best maximum grain size in the disk is mostly determined by the major axis profiles. To better understand the effect of grain size on the major axis profiles at our three wavelengths of interest, we present some models in Fig.~\ref{fig:maj_cut}. We chose to show models where only the maximum grain size varies, while the other parameters are fixed to $H_{ld, 100au}=3au$, $i=90^\circ$, $p=-1$, $\beta=1.14$ (the best values from Fig.~\ref{fig:integrated_grid}). The corresponding dust masses of each model (as determined in our first modeling step) are  $M_{dust} = 10^{-4} M_\odot$ for the models with $a_{max}=$100$\mu$m and $a_{max}=$1mm, $M_{dust} = 3\times10^{-4} M_\odot$ for the model with $a_{max}=$1cm. 

In Fig.~\ref{fig:maj_cut}, we find that the models with $a_{max} = 100\mu$m and $a_{max}= 1$mm can simultaneously reproduce the radial extent of the data at 0.9mm, 2.1mm, and 9.2mm, even though radial drift is not included in the modeling. On the contrary, the model with $a_{max}= 1$cm is unable to reproduce the brightness of the three bands simultaneously. Because of the presence of large grains, the radial profile of that model at 9.2mm is too radially extended to reproduce the data. Most combinations of ($H_{ld, 100au}$, $i$, $p$, $\beta$) also exclude a maximum grain size of $a_{max}= 1$cm as can be seen in Fig.~\ref{fig:integrated_grid}. In Appendix~\ref{appx:additional_models} we show a similar result when considering the $\bar{\chi}^2_{maj}$ values for different pairs of ($a_{max}$, $p$) and ($a_{max}$, $i$).  

This result indicates that if there are some grains of 1cm in the disk, they can not be well mixed with the gas and extending to the outer disk regions, because the emission would then appear too radially extended at 9.2mm. Such grains of 1cm might however be more concentrated radially than smaller grains, which can not be tested with the current data and modeling. Besides, the models show that if the maximum grain size is between $a_{max}=100\mu$m and $a_{max}= 1$mm, optical depths effects alone can be at the origin of the apparent radial segregation between millimeter and centimeter observations. 
This contrasts with previous radiative transfer modeling results of an other edge-on Class I source CB~26, who found a maximum grain size of 5cm throughout the disk~\citep{Zhang_2020}. 

\begin{figure*}
    \centering
    \includegraphics[width = 1\textwidth, trim={0.8cm 0cm 0.4cm 0.6cm}, clip]{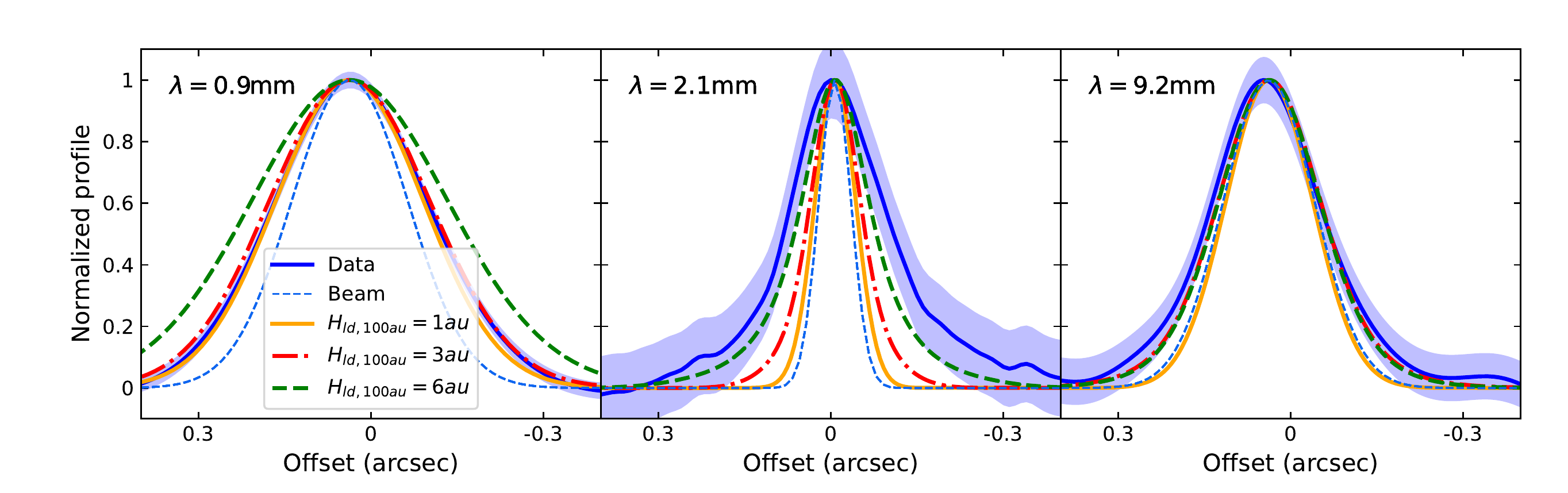}
    \caption{Effect of scale height on the minor axis profiles, for models with $a_{max} = 100\mu m$, $p=-1$, $\beta=1.14$, and $i=90^\circ$.}
    \label{fig:min_cut}
\end{figure*}

\subsubsection{Vertical extent of large dust particles}
\label{sec:mean_vertical}
Our grid models were performed for 5 different scale heights of the large dust grains. In Fig.~\ref{fig:min_cut}, we present the minor axis profiles of the data and models, integrated over the major axis extent of the disk (see Sect.~\ref{sec:fitting}). We represent models with $a_{max} = 100\mu m$, $p=-1$, $\beta=1.14$, $i=90^\circ$, and varying $H_{ld, 100au}$. The models with $H_{ld, 100au}=1$, 3 and 6au respectively have a dust mass of $M_{dust} = 3\times10^{-5}$, $ 10^{-4}$, and $ 3\times10^{-4} M_\odot$. Due to the high inclination of the models, these cuts are dominated by the vertical extent of the disk.

The ALMA 2.1mm image is the most resolved in the vertical direction~\citep{Villenave_2020}, it is thus the wavelength that is expected to be able to provide most constraints on the vertical extent of the disk. Indeed, we find that while the 1au and 3au models correctly match the respectively marginally resolved and unresolved minor axis profiles of the ALMA 0.9mm and VLA 9.2mm images, they appear to be too vertically thin to reproduce the 2.1mm images. In contrast, we find that the ALMA 0.9mm minor axis profile is not well reproduced by the models with scale height of 6au which appear too vertically extended to match the profiles. As shown in Fig.~\ref{fig:integrated_grid}, the best compromise between all bands is achieved for a scale height of 3\,au, while models with 1au or 6au of scale height can be excluded as they are unable to reproduce either the 2.1mm or the 0.9mm minor extent. In Appendix~\ref{appx:additional_models}, we show that these constraints on the large dust vertical extent are similar for the different inclinations and flaring exponents in our grid. In particular, we emphasize that while the models shown in Fig.~\ref{fig:min_cut} have an inclination of 90$^\circ$, our grid also explored less inclined disks and allowed us to exclude the possibility of IRAS04302 being both less inclined (88$^\circ$ of inclination) \emph{and} very settled (see Sect.~\ref{sec:grid_models}).\\

\begin{figure}
    \centering
    \includegraphics[width = 0.45\textwidth]{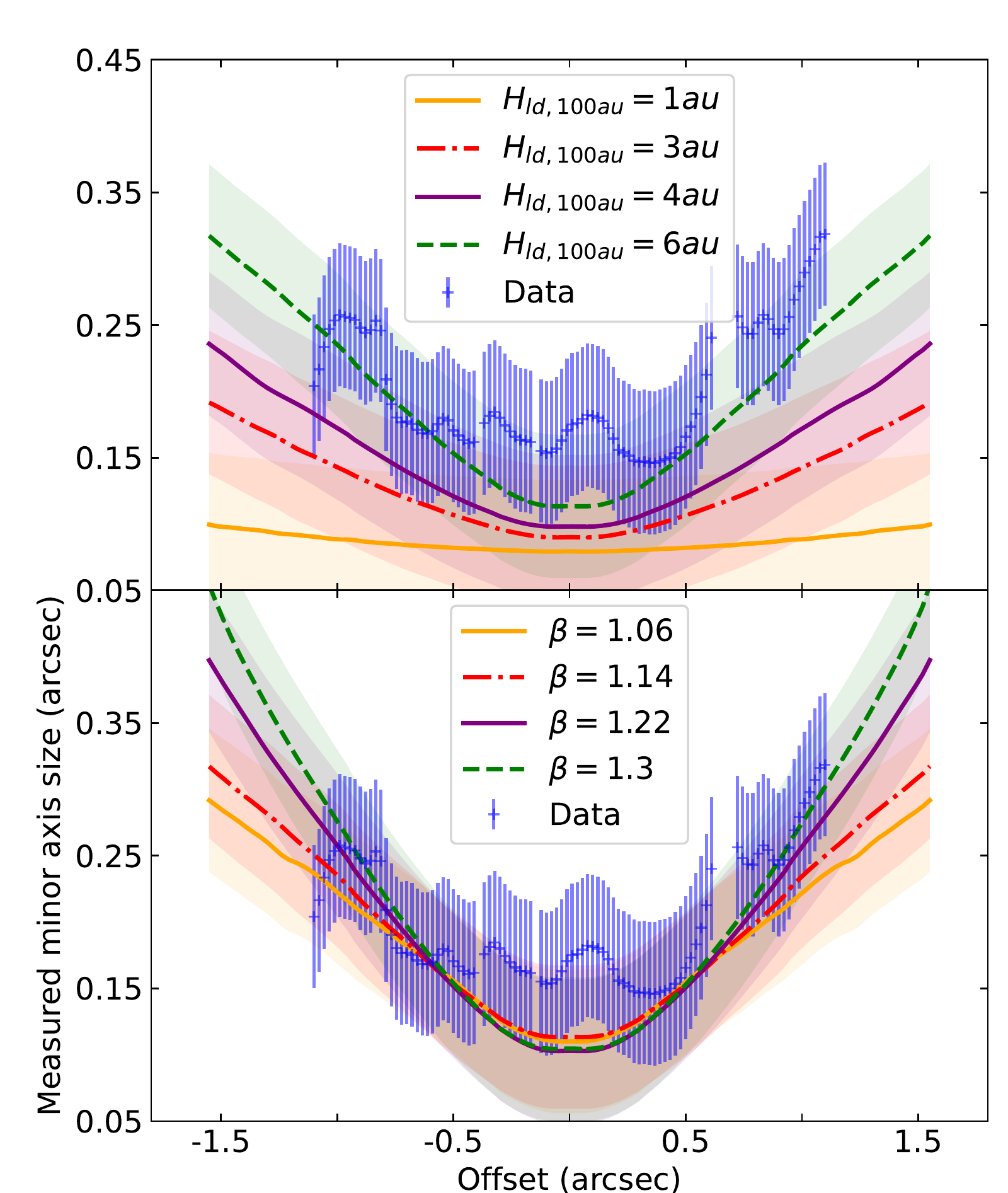}
    \caption{\emph{Top:} Minor axis size as a function of radius of the 2.1mm data and models for $\beta = 1.14$ and varying $H_{ld, 100au}$. \emph{Bottom:} Minor axis size as a function of radius of the 2.1mm data and models with $H_{ld, 100au} = 6$au, and varying $\beta$. Shaded regions show the beam size in the direction of the minor axis cuts. }
    \label{fig:radial_fwhm}
\end{figure}

Up to here, we focused on the minor axis cut averaged over the full disk. However, \citet{Villenave_2020} previously identified that the minor axis size of IRAS04302, measured at 2.1mm, increases with radius. Now, we present the impact of different parameters on the variation of the minor axis extent at different distances from the central star. Following \citet{Villenave_2020}, we fitted a Gaussian to the minor axis profiles obtained at a different locations. We plot the resulting full width half maximum (FWHM) in Fig.~\ref{fig:radial_fwhm} for two sets of models varying either $H_{ld, 100au}$ or $\beta$. In the two figures, we fix $i=90^\circ$, $a_{max}=100\mu$m, $p=-1$. Then the top figure shows the impact of the dust scale height on the minor axis size variation, for $\beta=1.14$, and the bottom figure shows the effect of flaring on the shape of the minor axis size variation, for $H_{ld, 100au} = 6$au\footnote{The models with varying flaring respectively have a dust mass of $M_{dust} = 3\times10^{-4}$ for $\beta = 1.06$ and 1.14, and $M_{dust} = 3 \times10^{-4} M_\odot$ for $\beta = 1.22$ and 1.3. Those with varying scale height are the same as in Fig.~\ref{fig:min_cut}, and the model with $H_{ld, 100au}=4$au has a dust mass of $M_{dust} = 2\times10^{-5} M_\odot$.}.

We find that while the $H_{ld, 100au}=1$au model does not show any clear variation of the minor axis size with respect to the distance to the star, every other model does. The model with a dust scale height of 6au provides the best match to the profile and absolute value. On the other hand, the bottom part of Fig.~\ref{fig:radial_fwhm} shows that the impact of the flaring exponent is less important than the impact of the scale height. For the range of flaring values explored within our grid and a dust scale height of 6au, all models are consistent with the data within the uncertainties, which does not permit to constrain the flaring of the large dust particles in this system.\\

To summarize, using a grid of radiative transfer models, we are able to constrain the scale height of the large grains in the Class I IRAS04302. We find that it is of a few astronomical units at a radius of 100au, where values lower than 1au and higher than 6au are excluded by our analysis. 
We obtain an upper limit consistent with the results of Lin et al. (in preparation), who modeled high angular resolution 1.3mm observations of IRAS04302, and also consistent with previous modeling results from \citet{Grafe_2013}, using vertically unresolved millimeter data. The flaring exponent $\beta$ of the millimeter disk could however not be constrained as its effect are minimal on the minor axis profile radial variation.

We note that no model is able to simultaneously reproduce the 0.9mm and 2.1mm data (see Fig.~\ref{fig:min_cut}). We suggest that this might be due to the simplifying assumptions that we made for our modeling, namely that we consider fully mixed grains without substructures, that we use a unique maximum grain size, and/or that we determine the dust mass by simultaneously reproducing the major axis profiles of the 0.9mm, 2.1mm, and 9.2mm observations.  
In particular, our results might suggest that the optical depth change between 0.9mm and 2.1mm is less than what is assumed in the models, such that the 0.9mm image could appear less vertically extended. Besides, the 0.9mm image is also  only marginally resolved in the vertical direction, such that further observations at higher angular resolution would allow to further constrain the vertical extent of large dust in IRAS04302. We note that Lin et al. (in preparation) estimated a dust scale height of about 6au at 100au when working on independent observations of IRAS04302 at 1.3mm with high angular resolution. Their constraints indeed appears more consistent with our results based on our most resolved observation (2.1mm) than from the constraints based on the less resolved 0.9mm image.

\section{Discussion}
\label{sec:discussion}
\subsection{Dust and gas vertical extent in IRAS04302}

In Sect.~\ref{sec:mean_vertical}, we found that the large dust scale height is constrained within $1\text{au}<H_{ld, 100au}<6\text{au}$. This result can be compared to the gas scale height that we estimate in Appendix~\ref{appx:match_thermal_SED}. We used the midplane temperature of our best model to which we added an envelope and a disk layer of small grains to obtain a better representation of the disk temperature structure. We obtained a midplane temperature of 21\,K at 100\,au from the central star\footnote{This temperature is in perfect agreement with the location of the CO snowline by \citet{vant_Hoff_2020} based on the disappearance of CO from the midplane at 100au~\citep[see also][]{Podio_2020}.}, which translates into a gas scale height of $H_{g, 100au}\sim 7$au, for a stellar mass of $M_\star =  1.6  M_\odot$ (Lin et al. in preparation).  The gas scale height appears to be slightly larger than our upper limit of the large dust scale height, suggesting that there is at least some modest amount of vertical settling occurring in the Class~I IRAS04302, in agreement with previous conclusions by \citet{Grafe_2013}.

From the difference of scale height between the gas and the millimeter dust grains, we can also estimate the degree of coupling of dust with gas, parameterized by the $\alpha/St$ ratio, where $St$ is the Stokes number and $\alpha$ the turbulence parameter. Indeed, assuming that turbulence balances gravity, for grains in the Epstein regime ($St\ll1$) and under the assumption of $z\ll H_g$, the dust scale height can be written as:
\begin{equation}
    H_d = H_g \left(1 + \frac{StSc}{\alpha}\right)^{-1/2}
\end{equation}
with $Sc$ the Schmidt number, the ratio between turbulent viscosity and turbulent diffusivity. Following previous studies \citep{Dullemond_2018, Rosotti_2020, Villenave_2022}, we assume $Sc=1$, which is valid for particles with $St\ll1$~(\citealt{Youdin_2007}, see also \citealt{Johansen_2005}). For $H_{ld, 100au}>1$au, and $H_g\sim 7$au, we find that $ [\alpha/St]_\mathrm{100au} >2\times 10^{-2}$. This value is higher than previous estimates on Class II systems (Table~\ref{tab:other_systems}), which we discuss in Sect.~\ref{sec:comparison}.

\begin{table*}[]
    \begin{center}
    \caption{Stellar mass, outer radius, bolometric luminosity, accretion rate, evolutionary stage, and $[\alpha/St]_{100au}$ for 7 disks with constraints on their large dust height from ALMA observations. }
    \label{tab:other_systems}
    \begin{tabular}{ccccccccl}
    \hline\hline
     & $M_\star$ & $R_{out,mm}$ & $L_{bol}$& Accretion  &Class & $[\alpha/St]_\mathrm{100au} $ & Dust Height & References\\
     &($M_\odot$)& (au)& ($L_\odot$)&($M_\odot/yr$)& \\
    \hline
        HH212 & 0.2-0.3 & 60 &9.0& 2-4$ \times 10^{-6}$ & 0 & $-$& Vertically thick& 1, 2, 3, 4\\ 
        L1527 & 0.2-0.4 & 54& 2.75 &5-11$\times 10^{-7}$&0/I & $-$& Vertically thick & 5, 6, 7, 8 \\
        VLA 1623 W& 0.45& 50 &0.13 & 2.5 $\times 10^{-8}$&0/I& $-$& Vertically thick & 9, 10, 11, 12 \\ 
        IRAS04302 & 1.6 & 300 &0.15-0.67 & 0.7-3$\times 10^{-8} $&I & $>2\times 10^{-2}$ &$1\text{au}<H_{ld, 100au} < 6\text{au}$ & This work, 13, 14 \\
        HL Tau & 1.7 & 150& 3.5-15 &$8.7 \times 10^{-8} $& I/II& $-$ &$H_{ld, 100au} < 1\text{au}$ & 14, 15, 16\\ 
        HD163296 &2.0 & 240 & $-$&$4.5\times10^{-7}$&II& $<1\times 10^{-2}$ & $H_{ld, 100au} < 1\text{au}$ & 17, 18, 19, 20, 21\\
        Oph163131 & 1.2 & 150& $-$ &$-$ & II& $<5\times 10^{-3}$ &$H_{ld, 100au} < 0.5\text{au}$ & 22, 23\\
    \hline
    \end{tabular}
\end{center}
\tablecomments{For HH212, L1527, VLA 1623 W, and IRAS04302, we recalculated the accretion rates based on their bolometric luminosity (see main text). For HD163296 and HL Tau, we report estimates based on the UV excess or HI recombination line luminosity. The outer radii and dust height correspond to the radial extent or dust height inferred for the dust detected at millimeter wavelengths with ALMA.}
\tablerefs{(1) \citet{Lee_2014}; (2) \citet{Codella_2014}; (3) \citet{Lee_2017}; (4) \citet{Lin_2021}; (5) \citet{Maret_2020}; (6) \citet{Tobin_2008}; (7) \citet{Sheehan_2022}; (8) \citet{Ohashi_2022}; (9) Mercimek et al. in prep; (10) \citet{Harris_2018}; (11) \citet{Murillo_2018}; (12) \citet{Michel_2022}; (13) Lin et al. in prep; (14) \citet{Robitaille_2007}; (15) \citet{Pinte_2016}; (16) \citet{Beck_2010}; (17) \citet{Teague_2019}; (18) \citet{Muro-Arena_2018}; (19) \citet{Mendigutia_2013}; (20) \citet{Doi_Kataoka_2021}; (21) \citet{Liu_2022}; (22) \citet{Flores_2021}; (23) \citet{Villenave_2022}}
\end{table*}

\subsection{Comparison with other systems}
\label{sec:comparison}

We now aim to look for any trend of variation of settling efficiency with different parameters, such as stellar mass, disk outer radius, and time. 
In Table~\ref{tab:other_systems}, we compiled various parameters for 7 disks of different evolutionary stage with constraints on large dust vertical height~\citep[HH212, L1527, VLA  1623 W, IRAS04302, HL Tau, HD163296, Oph163131,][]{Lin_2021, Sheehan_2022, Ohashi_2022, Natakani_2020, Michel_2022, Pinte_2016, Doi_Kataoka_2021, Liu_2022, Villenave_2022}. 

The stellar masses gathered in Table~\ref{tab:other_systems} are all dynamical estimates based on gas Keplerian rotation. For the Class I and II disks, the abundant $^{12}$CO lines were used, while rarer species such as C$^{17}$O, HCO$^+$ or SO were considered for the more embedded Class 0 protostars (see references in Table~\ref{tab:other_systems}). We also recalculated the mass accretion rates of the least evolved systems (HH212, L1527, VLA 1623 W, IRAS04302) based on their bolometric luminosity using  $\dot{M}_{acc} = \frac{L_{bol}R_\star}{GM_\star ( 1 - R_\star/R_{in})}$~\citep[][]{Gullbring_1998}, assuming that the bolometric luminosity is dominated by the accretion luminosity. We assume a stellar radius of $R_\star= 2 R_\odot$~\citep{Lee_2020}, and an inner radius of the accretion disk of $R_{in} = 5 R_\star$~\citep[][]{Gullbring_1998}. For HD163296 and HL Tau, we report estimates based on the UV excess or HI recombination line luminosity, and we note that no clear sign of accretion were found in the spectra of Oph163131~\citep{Flores_2021}. Finally, the last column of Table~\ref{tab:other_systems} reports the dust height of the different systems. For the most evolved sources, specific radiative transfer modeling were performed, allowing to quantify the vertical extent of large dust  particles and to compare it with that of the gas~\citep{Pinte_2016, Doi_Kataoka_2021, Liu_2022,  Villenave_2022, Wolff_2021}. On the other hand, for the younger Class~0 systems, determining the gas height is complex due to the presence of a dense envelope and the different papers did not provide quantitative measurement of the gas and dust scale height. Nevertheless, using radiative transfer, geometrical analysis in the uv-plane, or by measuring the apparent size in the image plane, the authors argued that the young disks of HH212, L1527, and VLA 1623~W are geometrically thick, and possibly not affected by vertical settling~\citep{Lin_2021, Sheehan_2022, Ohashi_2022, Michel_2022}.

From  Table~\ref{tab:other_systems}, we find that the millimeter dust height appears to correlate with the evolutionary class. In particular, the Class I IRAS04302 appears less settled than Class II and I/II disks (HD163296, Oph163131, HL Tau). In addition, while we identified at least moderate level of settling in IRAS04302, previous studies of younger Class~0  and 0/I systems (HH212, L1527, VLA1623W) suggested that settling did not occur at all in these young sources.  
Our results thus seems to support an evolution of settling efficiency with time.

In Table~\ref{tab:other_systems}, we also find that the younger sources have the least massive stars and least extended disks, which might impact the settling efficiency. To investigate the differences in dust height between the disks, we estimate the timescale for settling under simple assumptions. When only the thermal speed of molecules and vertical settling are considered, in other words in the absence of vertical turbulence and infalling material, the timescale for settling is given by~\citet{Armitage_2015}:
\begin{equation}
    t_\mathrm{settle} = \frac{\rho}{\rho_m}\frac{v_\mathrm{th}}{a}\frac{1}{\Omega_K^2} \propto \rho \frac{\sqrt{T}}{a}\frac{R^3}{M_\star}
    \label{eq:t_Set}
\end{equation}
where $\rho_m$ is the dust material density, $\rho$ the gas density, $a$ is the particle size, $v_{th} = \sqrt{8k_BT/\pi\mu m_H}$ is the thermal speed of the molecules, and $\Omega_K = \sqrt{GM_\star/R^3}$ the disk rotation frequency, which depends on the stellar mass ($M_\star$) and the local radius ($R$). The settling timescale strongly depends on the radius, particle size, stellar mass, and gas density. Specifically, settling is expected to be faster for larger particles, larger stellar mass, and at the inner regions of the disk. If they were in the same evolutionary stage, smaller disks, such as HH212, L1527, and VLA 1623 W, should thus appear more settled than larger disks, such as IRAS04302, HL Tau, HD163296, Oph163131, which we do not observe.

Using the generic numerical values given by \citet{Sheehan_2022}, $\rho = 10^{-12}$\,g\,cm$^{-3}$, $\rho_m=3$\,g\,cm$^{-3}$, and $T=50$\,K,  we find that the settling timescale for grains of 1mm is $t_{\mathrm{settle, 100au}}\sim0.4-0.09$~Myr at 100au, and $t_{\mathrm{settle, 50au}} \sim0.05-0.01$ Myr at 50au, for $M_\star=0.45-2M_\odot$. At 100au from the star, these estimates are comparable to the lifetime of Class~0 systems~(0.13-0.26 Myr, \citealt{Dunham_2015}; see also \citealt{Kristensen_2018}). This suggests that, if they are located at 100au from the star, 1mm grains do not have time to settle at the Class 0 stage, while at 50au they are expected to settle before the end of the Class~0 stage.   
On the other hand, Class~I objects have longer lifetimes~\citep[0.27-0.52 Myr,][]{Dunham_2015} and millimeter sized particles, if present, should already have totally settled both at 50au and 100au from the star.

Yet, we identify only a modest level of settling in IRAS04302, the only Class~I of Table~\ref{tab:other_systems}, and no evidence for settling in the small Class 0 and 0/I sources HH212, L1527, VLA 1623 W. This suggests that turbulence could be stronger in the earliest stages of star formation, preventing the dust grains to settle within the predicted timescale without turbulence. Interestingly, the values of $\alpha/St$ presented in Table~\ref{tab:other_systems} also indicate that, at 100au from the star, turbulence is stronger in the Class I IRAS04302 than in the Class II Oph163131 and HD163296. In addition, more evolved systems tend to have lower accretion rates (see Table~\ref{tab:other_systems}) which would also be consistent with evolution of turbulence with time, even though accretion rate does not necessarily relates to the turbulence at 100au. 

Alternatively, the difference in dust height between evolutionary classes might also be related to the difference in the environment external to the disk. Specifically, with an infalling envelope at the earliest stages of star formation, fresh grains are constantly being resupplied to the disk, both radially and vertically. So while the first grains to fall in may have had time to settle in the disk, recently added grains would not. Yet, recent studies showed that the grains falling onto the disk that are coming from the infalling envelope cannot have grown larger than a few 10s of microns, for timescale and density reasons \citep[e.g.,][]{Silsbee_2022}, such that it is not clear if this effect would be sufficient to explain the thickness of young disks when observed at millimeter wavelengths. 
Further studies are needed in order to determine which mechanism is dominant in setting the apparent height of disks at millimeter wavelengths for different evolutionary stages.

\section{Conclusions}
\label{sec:conclusion}

We presented new VLA observations of the Class I IRAS04302 disk, at 5 different wavelengths between 6.8mm and 66mm. The disk is detected both at 6.8mm and 9.2mm, while  processes other than dust continuum emission dominate the observations at 14mm, 40mm, and 66mm. We compare the disk size at centimeter wavelengths with its extent seen in millimeter ALMA observations at 0.9mm and 2.1mm and we find that the disk is about $ 70\%$ smaller at 9.2mm than at 0.9mm. The spectral index maps obtained with the different millimeter and centimeter observations allow us to identify an increase of the spectral index with radius, from $\sim$2 in the center to $\sim$3 in the outer disk, consistent with the size difference observed in the images. We also find that the spectral index obtained when including the centimeter observations is consistently higher than the spectral index using only the 0.9mm and 2.1mm images (by about 0.5dex except at the center of the disk), indicating that the observations at 9.2mm and 6.8mm are more optically thin than the ALMA observations.

With the goal of characterizing the distribution of large dust in IRAS04302, we produced a grid of radiative transfer models aiming to reproduce the 0.9mm, 2.1mm, and 9.2mm data. The models implement one layer of large grains without substructures, and we determine the dust mass for each combination of parameters based on the major axis profiles at the three wavelengths previously mentioned. We show in Appendix~\ref{appx:match_thermal_SED} that adding a small grain disk layer and envelope does not affect the shape of the millimeter/centimeter emission but allows to recover a more realistic temperature structure.

Our results indicate that if there are some grains of 1cm in the disk, they can not be well mixed with smaller dust particles and extend to the outer disk regions, because the emission would then appear too radially extended at 9.2mm. Such grains of 1cm might however be more concentrated radially than smaller grains, which can not be tested with the current data and modeling. Moreover, models with maximum grain sizes of 100$\mu$m or 1mm are able to reproduce the 0.9mm, 2.1mm, and 9.2mm major axis profiles simultaneously, which suggests that the apparent size difference between the millimeter and centimeter observations could be mostly due to optical depth effects. 
The models also favor a disk inclination very close to edge-on ($i>88^\circ$). However, no clear trend was found for the different values of the surface density exponent $p$ and the flaring exponent $\beta$ tested within our grid. 

Our modeling strategy also allows us to determine a plausible interval for the large dust scale height of $1\text{au}<H_{ld, 100au}<6\text{au}$. When compared to our estimate of the gas scale height $H_{g, 100au}\sim 7$au, based on the midplane temperature profile of our best model, we identify that the millimeter dust in the disk is subject to at least moderate vertical settling. The level of settling in this Class I disk contrasts with previous results of Class 0 systems, where no settling is occurring, and of Class II disks, that are found to be thinner. By estimating the timescale of vertical settling for millimeter sized particles and comparing it to the lifetime of Class~0 and Class~I systems, we suggest that this variation of settling efficiency with time is linked to some variation of turbulence or of external conditions with time. 

\begin{acknowledgments}
\emph{Acknowledgements:}
MV research was supported by an appointment to the NASA Postdoctoral Program at the (NASA Jet Propulsion Laboratory), administered by Oak Ridge Associated Universities under contract with NASA. EB acknowledges the Deutsche Forschungsgemeinschaft (DFG, German Research Foundation) under Germany´s Excellence Strategy – EXC 2094 – 390783311. CC-G acknowledges support by UNAM DGAPA-PAPIIT grant IG101321 and CONACyT Ciencia de Frontera grant number 86372. GD acknowledges support from NASA grants 80NSSC18K0442 as well as NNX15AC89G and NNX15AD95G/NExSS.
CC and LP acknowledge the EC H2020 research and innovation
programme for the project "Astro-Chemical Origins” (ACO, No 811312) and
the PRIN-MUR 2020 MUR BEYOND-2p (Astrochemistry beyond the second period elements, Prot. 2020AFB3FX).
This paper makes use of the following ALMA data: ADS/JAO.ALMA\#2016.1.00460.S. ALMA is a partnership of ESO (representing its member states), NSF (USA) and NINS (Japan), together with NRC (Canada), MOST and ASIAA (Taiwan), and KASI (Republic of Korea), in cooperation with the Republic of Chile. The Joint ALMA Observatory is operated by ESO, AUI/NRAO, and NAOJ.

\end{acknowledgments}

\vspace{5mm}
\facilities{ALMA, VLA}
\software{\texttt{CASA}~\citep{CASA_2022}, \texttt{mcfost}~\citep{Pinte_2006, Pinte_2009}, \texttt{Matplotlib}~\citep{Hunter_2007}, \texttt{Numpy}~\citep{Harris_2020}. }

\appendix

\section{Correction for processes other than dust emission}
\label{app:free-free}

\begin{figure*}
    \centering
    \includegraphics[width = 1\textwidth]{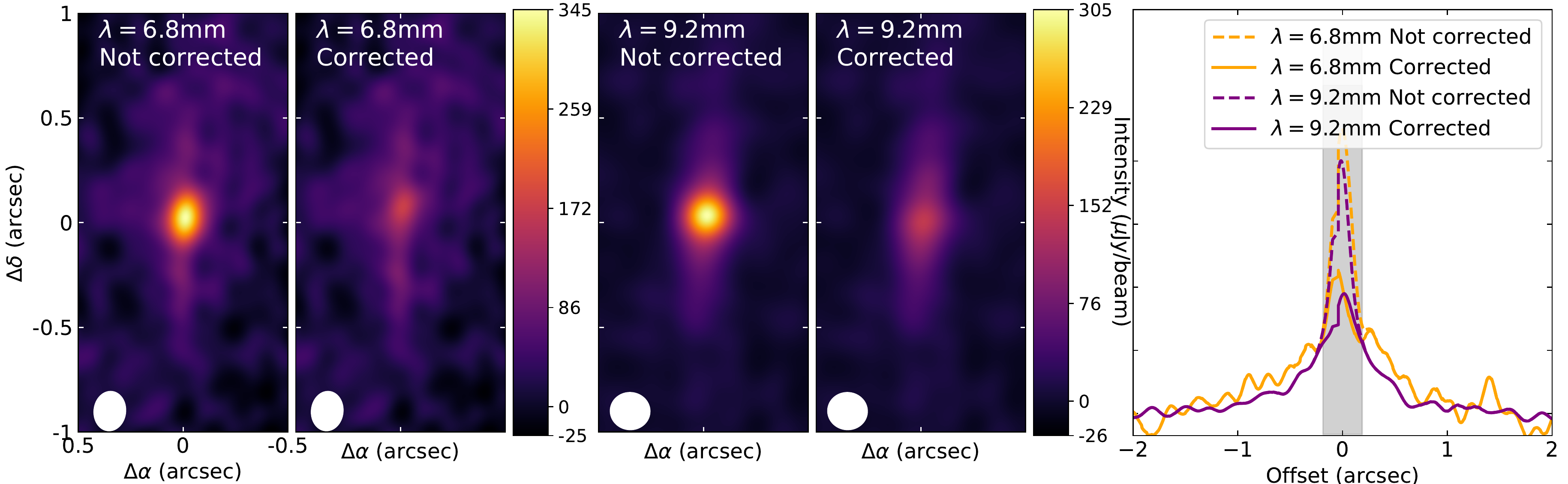}
    \caption{Resulting 6.8mm and 9.2mm images before (left panels) and after (middle panels) correction for non-disk emission, at the same angular and brightness scale. The right panel shows the major axis cut to the same brightness scale, illustrating that this correction only affects the central beam of the emission (marked with the filled grey region). }
    \label{fig:free-free_correction}
\end{figure*}

The variety of VLA observations allows us to perform a detailed analysis to remove  emission from processes other than dust thermal emission (also referred to as non-dust or wind/coronal emission throughout the text) present at the center of the disk. 
To do so, we follow the approach highlighted in \citet{Carrasco-Gonzalez_2019}. We first aligned the observations at the different wavelengths, using the CASA task \texttt{fixplanets}. Then, we produced an image combining the 66mm, 40mm, 14mm, and 6.8mm data together, aiming to model the non-dust emission of the disk. We use the \texttt{tclean} task with \texttt{nterm=2}, which allows us to obtain both the flux and spectral index of the emission. We also assume a briggs robust parameter of 0 in order to be sensitive only to the brightest emission of the maps.

We note that at 6.8mm and 9.2mm it is possible that both dust and non-dust emission are present in the inner regions. To determine whether to use the 6.8mm or 9.2mm observations to model the wind/coronal emission, we looked at the shape of the emission when imaged with a robust briggs parameter of 0. First, we note that this parameter value was chosen to decrease the sensitivity of the imaging, such that only the brightest emission (expected to be from processes other than dust thermal emission) will be visible in the final image. In addition, we expect wind/coronal emission to be concentrated within one beam, while dust emission from the disk should be more extended. We find that using a briggs 0 parameter the 6.8mm emission is concentrated within a central point source, and is thus likely dominated by processes other than dust thermal emission. On the contrary,  with a briggs parameter of 0, the 9.2mm image is extended, meaning that it still contains substantial dust emission. Thus, we include the 6.8mm but not the 9.2mm observations in the non-dust model.

From the combined 66mm, 40mm, 14mm, and 6.8mm image, we find that the wind/coronal emission is unresolved, as expected. We obtain a model consisting of only one point, located at the center of the disk, and with a flux density of 150$\mu$Jy at 26 GHz, and a spectral index of +0.46, consistent with moderately optically thick free-free emission~\citep{Rodmann_2006}. This allows to predict the contribution from processes other than dust thermal emission at 9.2mm and 6.8mm, assuming that such emission did not vary significantly between October and December 2021.

To remove coronal/wind contamination from the 6.8mm and 9.2mm image, we then inserted the non-dust model images (combined 66mm, 40mm, 14mm, and 6.8mm image) into the 'model' column of the 6.8mm and 9.2mm visibilities using the \texttt{ft} task, and subtracted the model column from the data column using \texttt{uvsub}. Finally, we produced a model image using the parameters presented in Sect~\ref{sec:VLA_Ka}. We present the 6.8mm and 9.2mm images and major axis cut before and after correction in Fig.~\ref{fig:free-free_correction}. These images clearly show that the correction significantly reduces the centrally peaked feature within 1 beam from the center, while keeping the outer regions unaffected. When modeling the disk (Sect.~\ref{sec:model}) we use the corrected 9.2mm image, but consider only the regions outside of 1 beam size to limit the effect of potential variability of wind/coronal emission.

\section{Simultaneous match of the thermal part of the SED and ALMA/VLA maps}
\label{appx:match_thermal_SED}

Previous observations and modeling~\citep{Lucas_Roche_1997, Padgett_1999, Wolf_2003} found that the scattered light is dominated by an envelope. In Sect.~\ref{sec:model}, we have however omitted this part in order to save some computer power in the generation of our grid. In this Appendix, we show that including a disk layer of small grains and an envelope does not significantly affect the shape of the 0.89mm, 2.1mm, and 9.2mm observations.  Our goal here is to produce a more realistic temperature structure of the disk that allows to match the millimeter fluxes and the thermal part of the Spectral Energy Distribution (SED, $\lambda \gtrsim20 \mu$m).

In order to obtain a better temperature structure of our disk, we thus add a disk layer of small grains and an envelope to the layer of large grains obtained as best model from our fitting methodology. We first added a layer of small grains with the parameters indicated in Table~\ref{tab:small_grain_gas}.
Then, we tested a small number of envelope parameters. For simplicity, we assume that the envelope is spherically symmetric and that the density follows a power law in radius: $\rho \propto r^\gamma$. Following \citet{Wolf_2003}, we also include a cavity in our modeling. We assume that no dust is present above a surface represented by $z = z_0 (r/r_0)^{\beta_{env}}$. We set the inner radius to 0.1au, the outer radius to 1000au, and varied the following parameters: $a_{max, env} \in [0.5, 1]$, $\gamma \in [-1, -2]$. $z_0 \in [1, 3, 5]$, and $M_{dust, env} \in [10^{-5}, 10^{-4}, 10^{-3}] $. Finally, we determined the best set of parameters based on a $\chi_{SED}^2$ estimate, using the SED match between 20 and 10000$\mu$m (7 points), and $\phi^2$ of the 0.9mm observations only.

\begin{table}[]
    \begin{centering}
    \caption{Small grains layer and envelope parameters.}
    \label{tab:small_grain_gas}
    \begin{tabular}{lcc}
    \hline\hline
    &&\emph{Envelope \& Cavity}\\
    $a_{min} - a_{max}$ &($\mu m$)& $0.005 - 0.5$\\
    $R_{in} - R_{out}$ &(au) & $0.1 -1000$ \\
    $M_{dust}$ &($M_\odot$)  & $1\cdot 10^{-5}$\\
    $\gamma$& ($-$)& -2 \\
    $z_0$, $r_0$, $\beta_{env}$ & (au, au, $-$) & 5, 1, 1\\
    \hline
    && \emph{Small grains disk layer}\\
    $a_{min} - a_{max}$ &($\mu m$)& $0.005 - 10$\\
    $Rin - Rout$ &(au) & $0.1 -300$ \\
    $M_{dust}$& ($M_\odot$)  & $5\cdot 10^{-7}$\\
    $H_{sd,100au}$, $\beta$, $p$ & (au, $-$, $-$) & 6.7, 1.14, -1 \\
    \hline
    \end{tabular}
    \end{centering}
    \tablecomments{The parameters for the small grain and envelope regions allow to produce a more realistic temperature structure of the disk than in Sect~\ref{sec:model}, but we emphasize that they are not expected to reproduce the previous scattered light observations and should not be considered as constraints on the envelope mass and shape.}

\end{table}
    
\begin{figure*}
    \centering
    \includegraphics[width = 1\textwidth, trim={0.8cm 0cm 0.4cm 0.6cm}, clip]{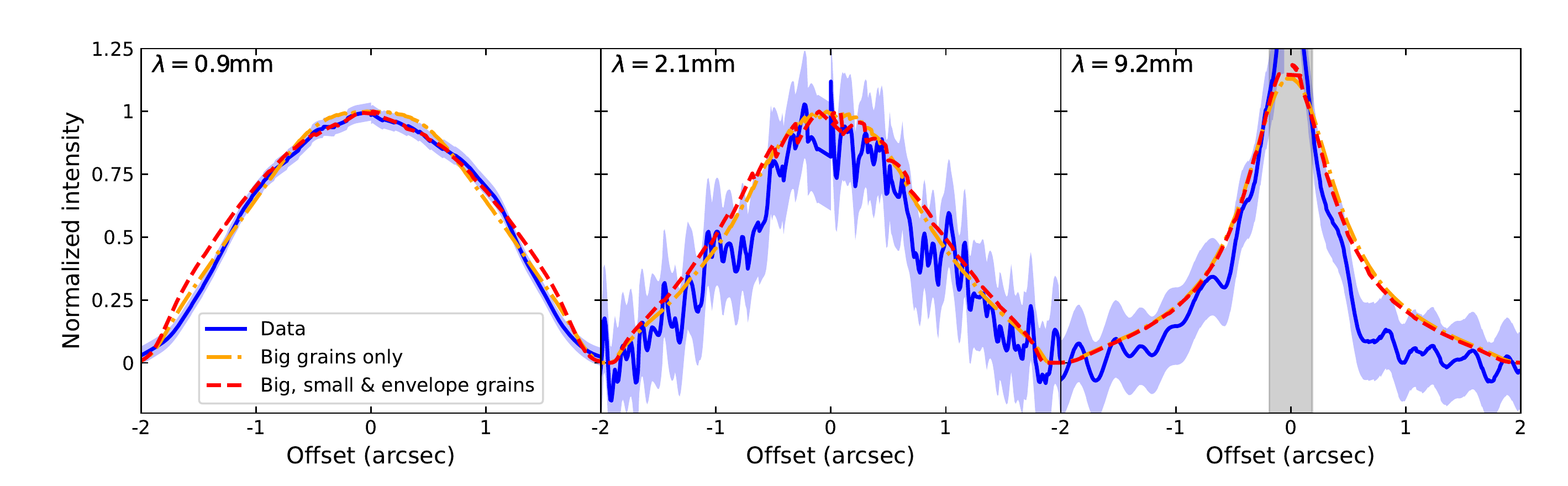}
    \includegraphics[width = 1\textwidth, trim={0.8cm 0cm 0.4cm 0.6cm}, clip]{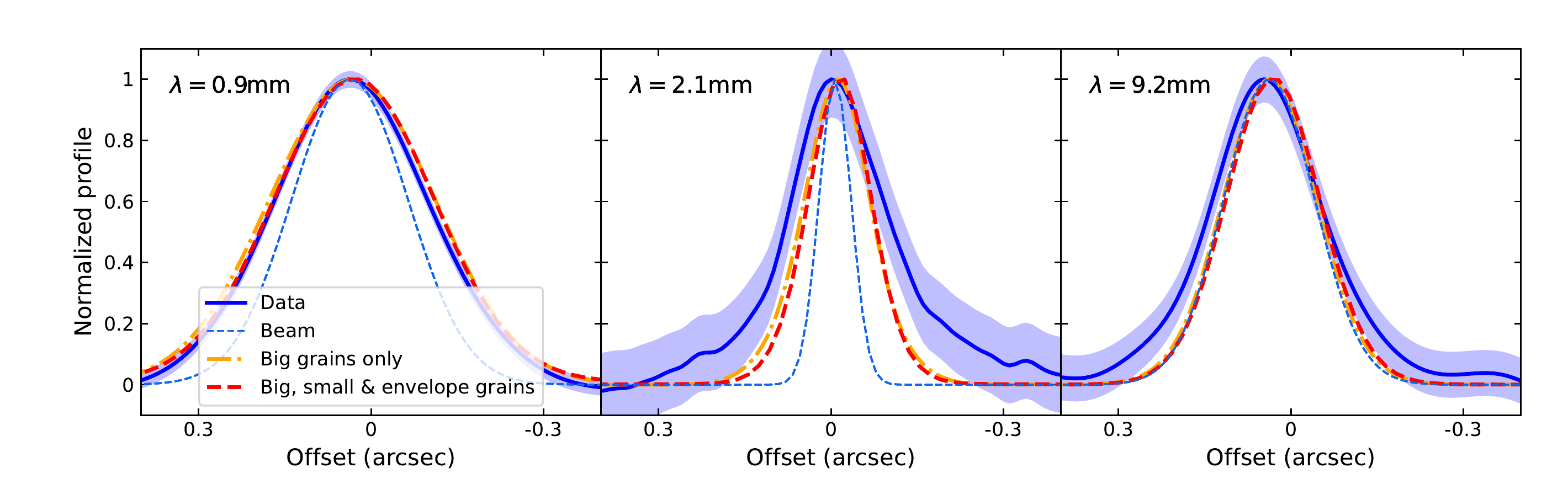}
    \caption{Comparison of the major axis profiles (top panels) and minor axis profiles (bottom panels) of the best model from Sect.~\ref{sec:grid_models} (in orange), that model plus an envelope and small grain layer (Table~\ref{tab:small_grain_gas}, in red), and the data (blue).}
    \label{fig:envelope}
\end{figure*}    
    
\begin{figure}
    \centering
    \includegraphics[width = 0.5\textwidth]{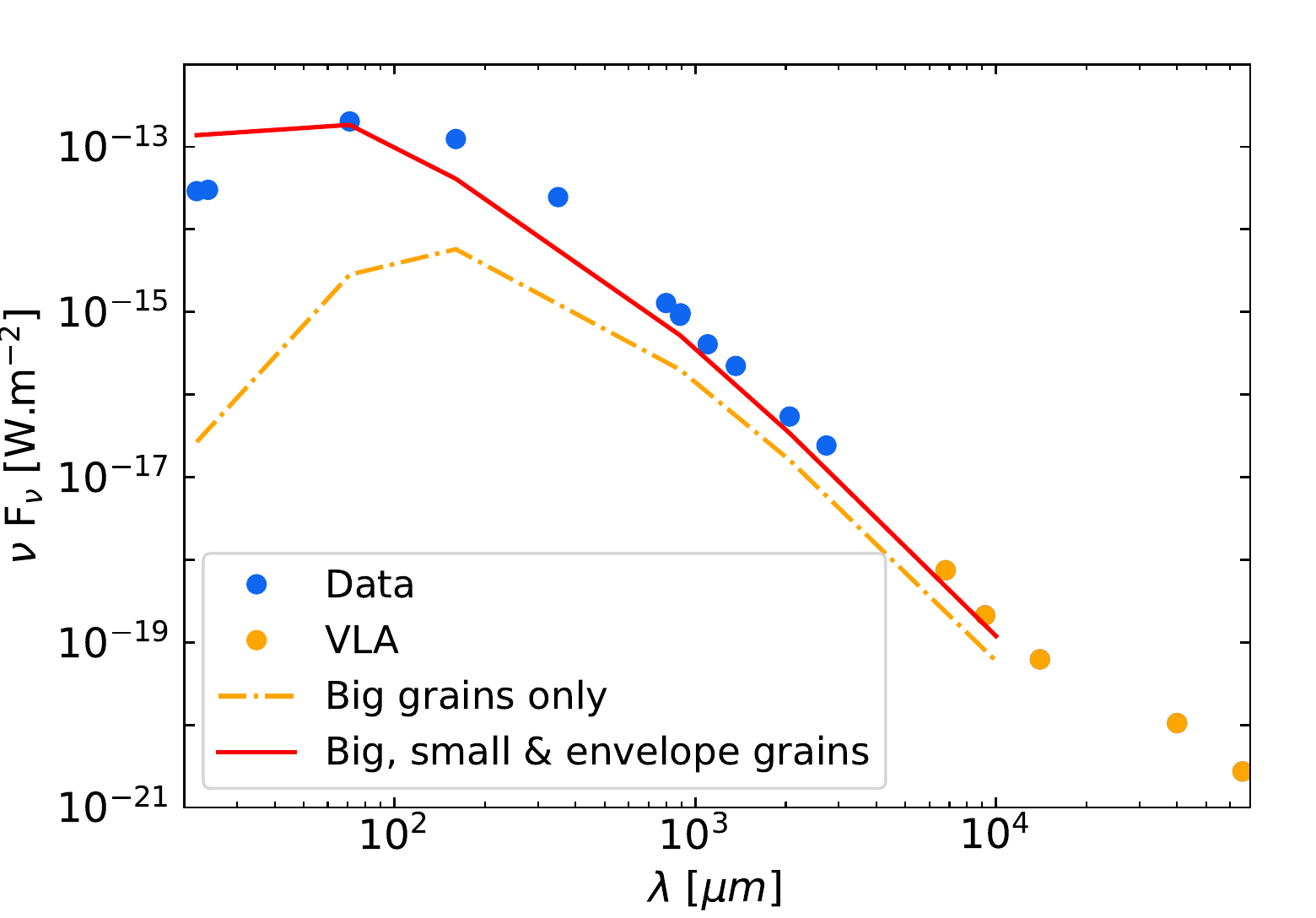}
    \caption{Comparison of the SED of the best model from Sect.~\ref{sec:grid_models} (in orange), that model plus an envelope and small grain disk layer (Table~\ref{tab:small_grain_gas}, in red), and the data (blue). The 6.8mm and 9.2mm VLA fluxes were corrected from processes other than dust emission. }
    \label{fig:envelope_sed}
\end{figure}

In Table~\ref{tab:small_grain_gas}, we show the resulting parameters for the small grain and envelope layers. We also include the comparison of the best big grains model ($H_d = 3 $au, $i=89^\circ$, $a_{max}=1000\mu$m, $\beta$=1.14, $p=-1.5$, and $M_d=0.0003 M_\odot$) with or without including small grains and an envelope in Fig.~\ref{fig:envelope} for the major and minor axis cut, and in Fig.~\ref{fig:envelope_sed} for the SED. As expected, we find that the thermal emission part of the model SED reproduces significantly better the data when a small grains and envelope layers are included. On the other hand, no significant differences are found in the shape of the millimeter/centimeter emission between the two models. Most of the differences lie in the outer edges of the major axis profile at 0.9mm.
This shows that, while the small grain and envelope layers are mostly invisible in the shape of the millimeter/centimeter observations, they contribute to obtain a more realistic disk structure and a larger total mm/cm flux, by allowing the midplane to be warmer. 
We emphasize that the parameters for the small grain and envelope region are not expected to reproduce the previous scattered light observations and should not be considered as constraints on the envelope mass and shape.\\

We find that adding the small grain and envelope layers allow to significantly increase the midplane temperature. At 100au from the central star, it goes from 11K with large grains only, to 21K when including the other layers. Using this midplane temperature value, it is possible to infer what is the gas scale height assuming that it is at the hydrostatic equilibrium, following:

\begin{equation}
    H(r) = \sqrt{\frac{k_B T(r) r^3}{G M_\star \mu m_p}}
\end{equation}
where $k_B$ is the Boltzmann constant, $\mu=2.3$ the reduced mass, $m_p$ the proton mass, $G$ the gravitational constant. Using stellar mass of $M_\star =$  1.6  M$_\odot$ (Lin et al. in preparation), we find that the gas scale height of this disk is $H_{g, 100au} \sim 7$ au. 

\section{Additional model comparisons}
\label{appx:additional_models}

\subsection{Residual maps of the best model}

\begin{figure}
    \centering
     \includegraphics[width = 0.5\textwidth]{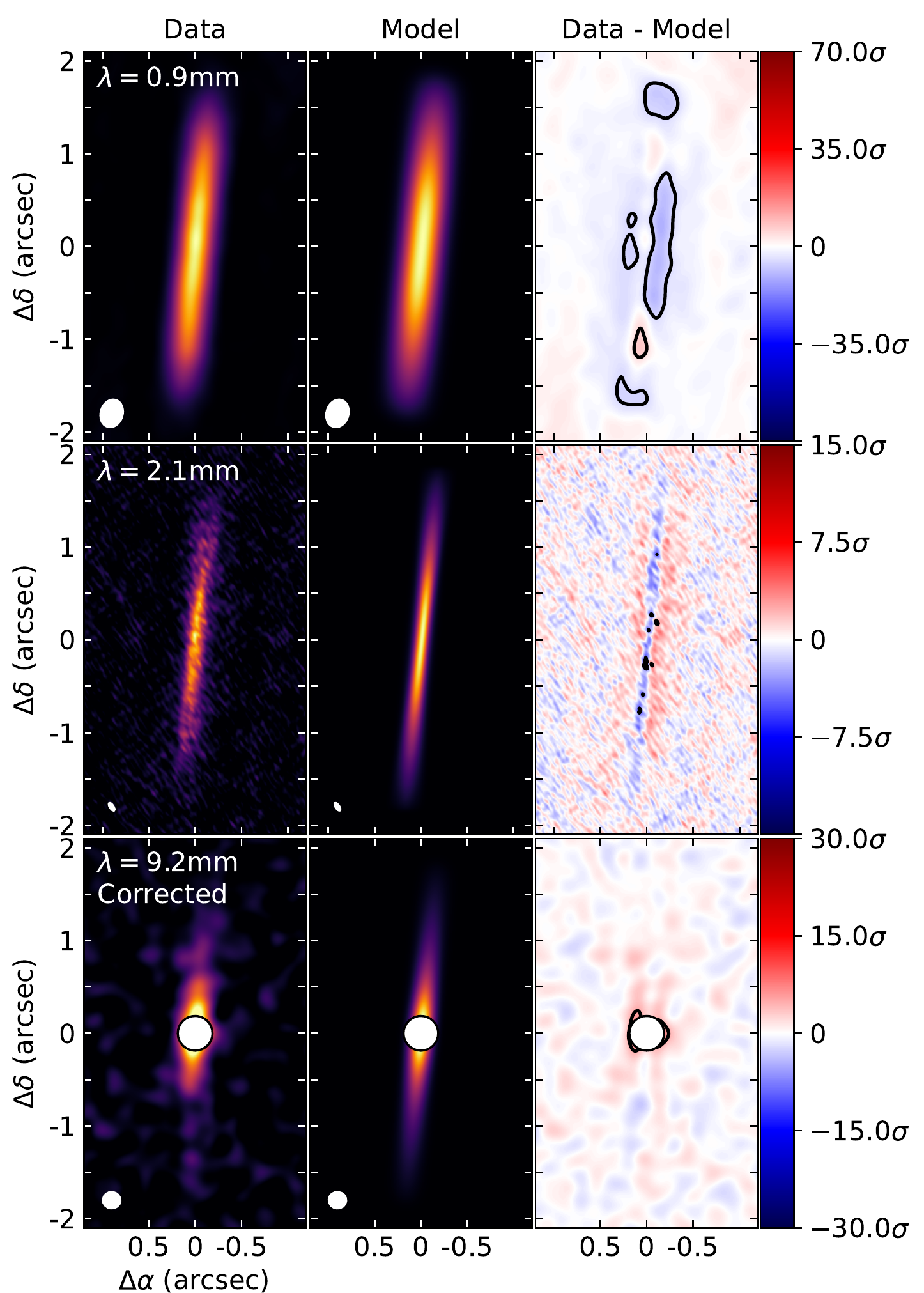}
    \caption{Left: Data images, Middle: Images of the best model, Right: Residual maps, with deviations by more than 5$\sigma$ indicated by black contours. The central mask on the 9.2mm maps indicate the region potentially affected by coronal/wind emission that is not considered in our modeling.}
    \label{fig:residuals}
\end{figure}

We present the residual maps of the best model in Fig.~\ref{fig:residuals}. The parameters for the best model are: $H_{ld, 100au} = 3 $au, $i=89^\circ$, $a_{max}=1$mm, $\beta$=1.14, $p=1.5$, and $M_{dust}=3\times 10^{-4} M_\odot$. We also note that the maps shown here do not have the small grain disk and envelope layer discussed in Appendix~\ref{appx:match_thermal_SED}.

\subsection{Effect of $p$ and $i$ on $a_{max}$}
\begin{figure}
    \centering
     \includegraphics[width = 0.5\textwidth]{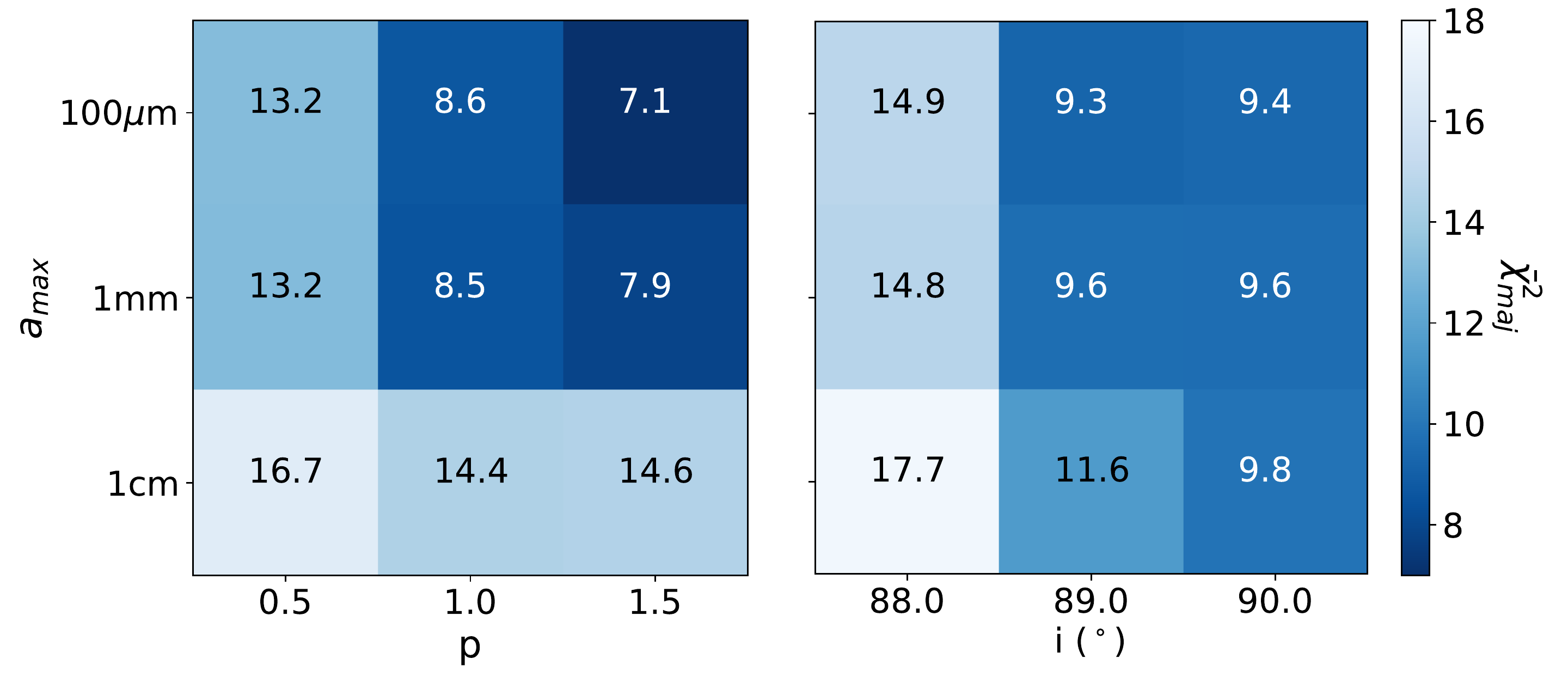}
    \caption{Median $\bar{\chi}_{maj}^2$ values for each combinations of ($a_{max}, p$) and ($a_{max}, i$). Darker colors indicate smaller $\bar{\chi}_{maj}^2$, and thus a better combination of parameters.}
    \label{fig:amax_i_p}
\end{figure}

While the maximum grain size has a clear impact on the shapes of the major axis profiles (Sect.~\ref{sec:major_profiles}, Fig.~\ref{fig:maj_cut}) it is possibly degenerated with three other parameters: $M_{dust}$, $i$, and $p$. In our first fitting step, we determine the best mass that allows to reproduce the major axis profile, thus the possible degeneracies between $M_{dust}$ and $a_{max}$ are already taken into account. In Fig.~\ref{fig:amax_i_p}, we show the detailed $\bar{\chi}_{maj}^2$ median matrix for each combination of ($a_{max}, i$) and ($a_{max}, p$). Both matrices show that independently of the value of $i$ and $p$, $a_{max}= 1$cm (lowest line) leads to significantly higher median $\bar{\chi}_{maj}^2$ values than lower maximum grain sizes. As indicated previously this implies that the maximum grain size in the system can not be 1cm in the outer regions of the disk and must be smaller.

\begin{figure*}
    \centering
    \includegraphics[width = 0.95\textwidth, trim={0.8cm 0cm 0.4cm 0.6cm}, clip]{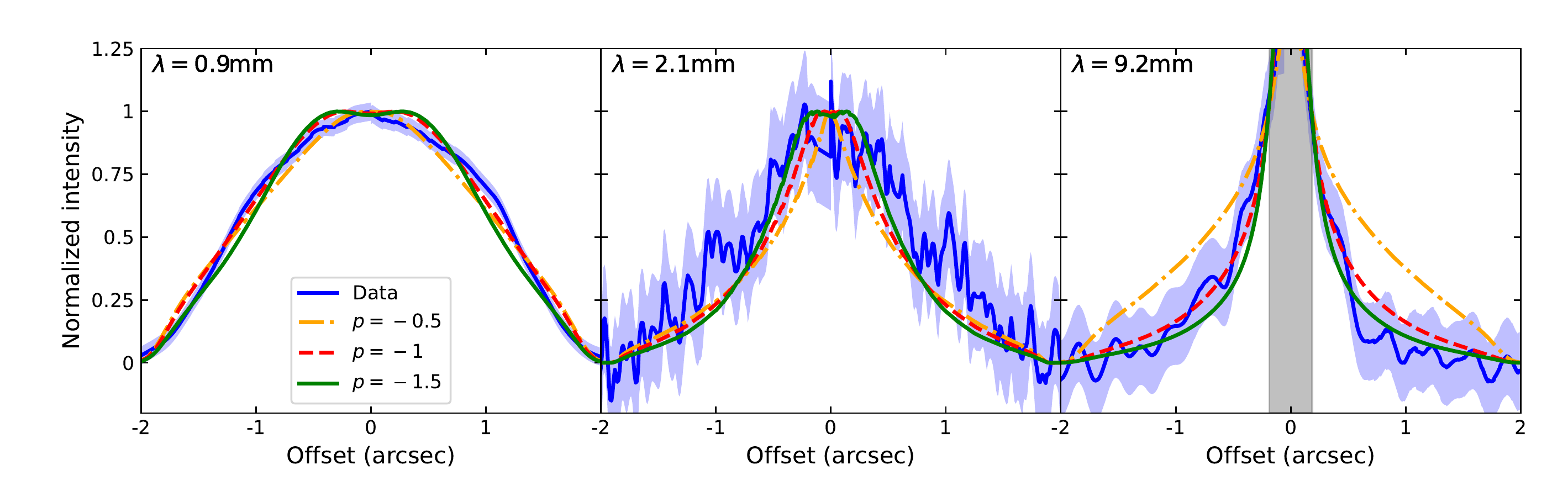}
    \includegraphics[width = 0.95\textwidth, trim={0.8cm 0cm 0.4cm 0.6cm}, clip]{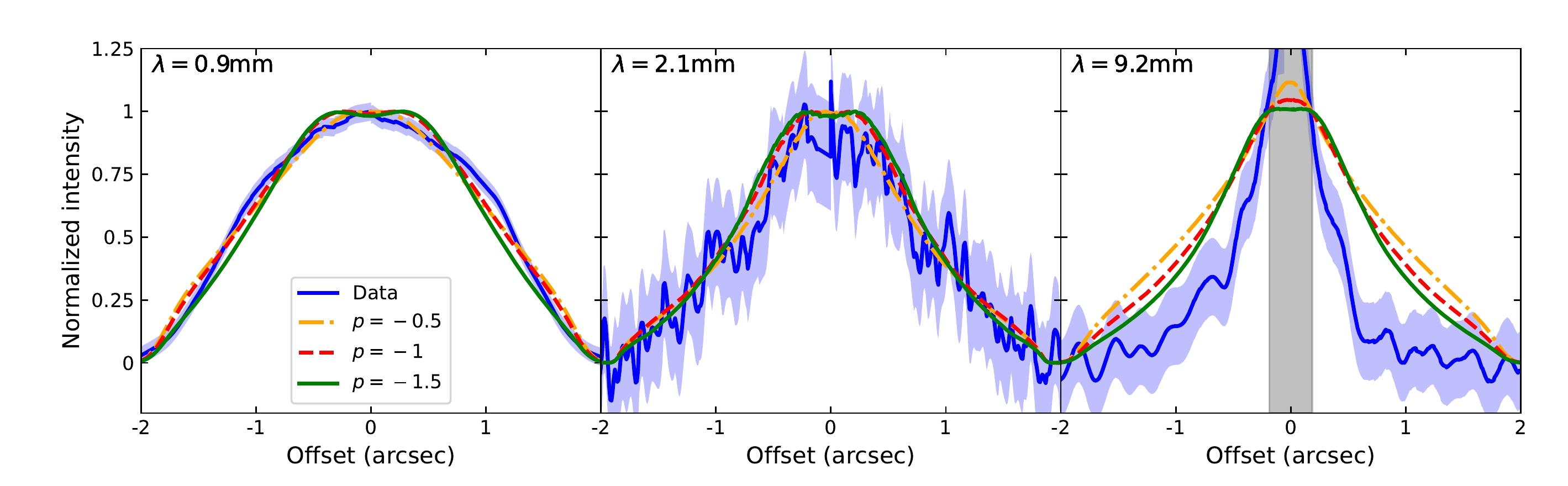}
    \caption{Effect of $p$ on the major axis profiles, for $a_{max} = 100\mu$m (top), and $a_{max}=1$cm (bottom).}
    \label{fig:maj_cut_p}
\end{figure*}

For all $a_{max}$ it is also clear that higher inclinations and higher surface density exponents are preferred ($i>88^\circ$, and $p>0.5$). This is because inclinations of 88$^\circ$ lead to models that are too centrally peaked. On the contrary, surface density exponent of $p=0.5$ implies a shallower radial profile, and leads the models to be too bright and radially extended at 9.2mm compared to the data. 
In Fig.~\ref{fig:maj_cut_p}, we show the effect of the surface density exponent on the shape of the major axis profiles, for a grain size of $a_{max} = 100\mu$m, and $a_{max}=1$cm. For the lower maximum grain size ($a_{max} = 100\mu$m), low values of surface density exponents are more strongly excluded than when  $a_{max} = 1$cm.

\subsection{Effect of $\beta$ and $i$ on $H_{ld, 100au}$}
\begin{figure}
    \centering
    \includegraphics[width = 0.5\textwidth]{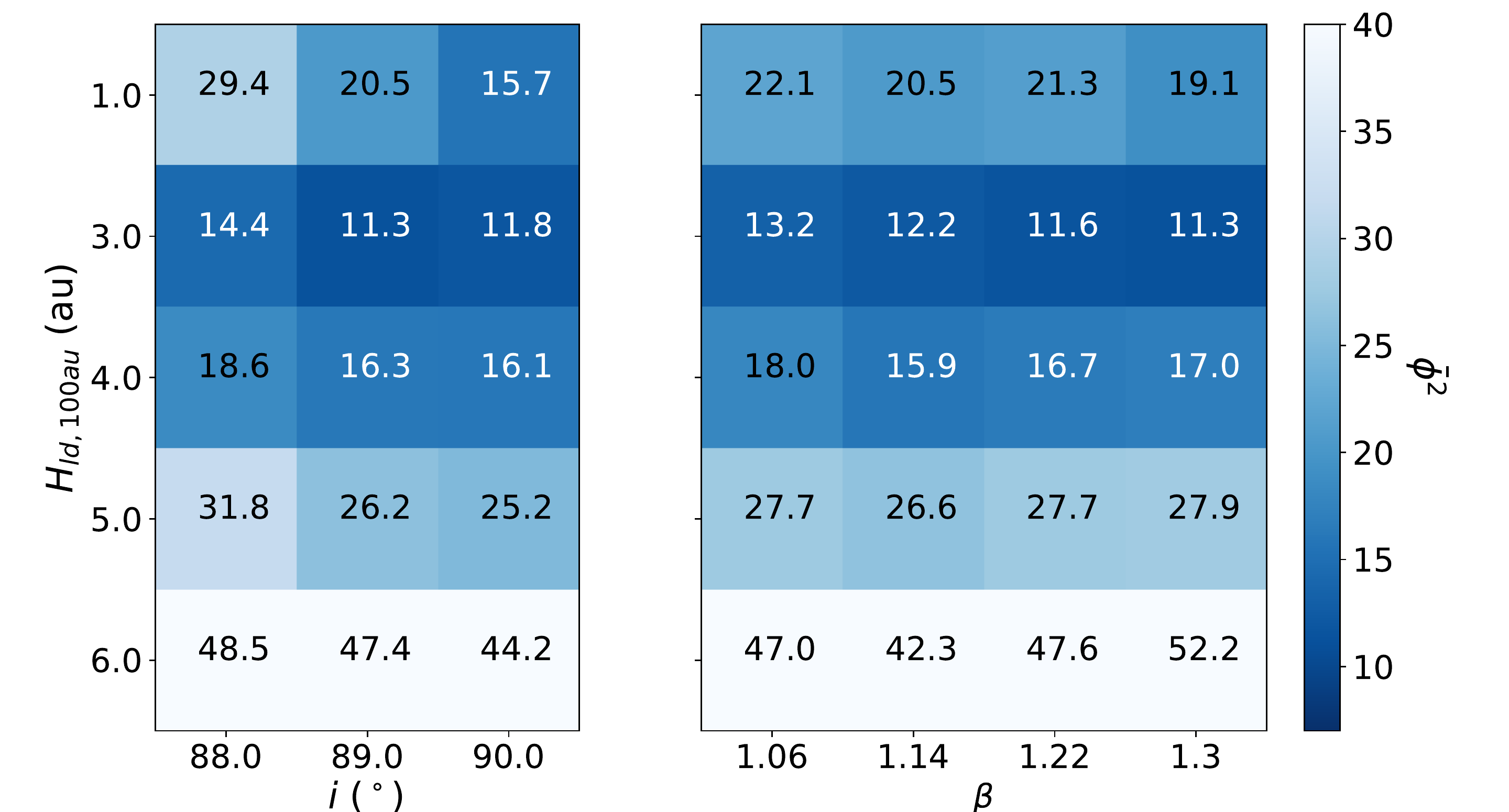}
    \caption{Median $\bar{\phi}^2$ for each combinations of ($H_d, i$) and ($H_d, \beta$). Darker colors indicate smaller $\bar{\phi}^2$ and thus a better combination of parameters. }
    \label{fig:hd_i_b}
\end{figure}

The models presented in Fig.~\ref{fig:min_cut} were computed for an inclination of $i=90^\circ$ and flaring of $\beta=1.14$. However, both the inclination and the flaring exponent $i$ and $\beta$ can also impact the apparent minor axis size of the disk. In Fig.~\ref{fig:hd_i_b}, we present the median $\phi^2$  matrix for each combinations of ($H_{ld, 100au}, i$) and ($H_{ld, 100au}, \beta$). These maps ($\phi^2$) show that larger inclinations and flaring are favored. Independently of the inclination and flaring exponent, we also find that a scale height of $H_{ld, 100au}\sim 3$au is favored.

\bibliography{biblio}{}

\begin{thebibliography}{}
\expandafter\ifx\csname natexlab\endcsname\relax\def\natexlab#1{#1}\fi
\providecommand{\url}[1]{\href{#1}{#1}}
\providecommand{\dodoi}[1]{doi:~\href{http://doi.org/#1}{\nolinkurl{#1}}}
\providecommand{\doeprint}[1]{\href{http://ascl.net/#1}{\nolinkurl{http://ascl.net/#1}}}
\providecommand{\doarXiv}[1]{\href{https://arxiv.org/abs/#1}{\nolinkurl{https://arxiv.org/abs/#1}}}

\bibitem[{{Andre} {et~al.}(2000){Andre}, {Ward-Thompson}, \&
  {Barsony}}]{Andre_2000}
{Andre}, P., {Ward-Thompson}, D., \& {Barsony}, M. 2000, in Protostars and
  Planets IV, ed. V.~{Mannings}, A.~P. {Boss}, \& S.~S. {Russell}, 59.
\newblock \doarXiv{astro-ph/9903284}

\bibitem[{{Andrews} {et~al.}(2018){Andrews}, {Huang}, {P{\'e}rez}, {Isella},
  {Dullemond}, {Kurtovic}, {Guzm{\'a}n}, {Carpenter}, {Wilner}, {Zhang}, {Zhu},
  {Birnstiel}, {Bai}, {Benisty}, {Hughes}, {{\"O}berg}, \&
  {Ricci}}]{Andrews_2018}
{Andrews}, S.~M., {Huang}, J., {P{\'e}rez}, L.~M., {et~al.} 2018, \apjl, 869,
  L41, \dodoi{10.3847/2041-8213/aaf741}

\bibitem[{{Ansdell} {et~al.}(2018){Ansdell}, {Williams}, {Trapman}, {van
  Terwisga}, {Facchini}, {Manara}, {van der Marel}, {Miotello}, {Tazzari},
  {Hogerheijde}, {Guidi}, {Testi}, \& {van Dishoeck}}]{Ansdell_2018}
{Ansdell}, M., {Williams}, J.~P., {Trapman}, L., {et~al.} 2018, \apj, 859, 21,
  \dodoi{10.3847/1538-4357/aab890}

\bibitem[{{Armitage}(2015)}]{Armitage_2015}
{Armitage}, P.~J. 2015, arXiv e-prints, arXiv:1509.06382.
\newblock \doarXiv{1509.06382}

\bibitem[{{Barri{\`e}re-Fouchet} {et~al.}(2005){Barri{\`e}re-Fouchet},
  {Gonzalez}, {Murray}, {Humble}, \& {Maddison}}]{Barriere-Fouchet_2005}
{Barri{\`e}re-Fouchet}, L., {Gonzalez}, J.~F., {Murray}, J.~R., {Humble},
  R.~J., \& {Maddison}, S.~T. 2005, \aap, 443, 185,
  \dodoi{10.1051/0004-6361:20042249}

\bibitem[{{Beck} {et~al.}(2010){Beck}, {Bary}, \& {McGregor}}]{Beck_2010}
{Beck}, T.~L., {Bary}, J.~S., \& {McGregor}, P.~J. 2010, \apj, 722, 1360,
  \dodoi{10.1088/0004-637X/722/2/1360}

\bibitem[{{Carrasco-Gonz{\'a}lez} {et~al.}(2019){Carrasco-Gonz{\'a}lez},
  {Sierra}, {Flock}, {Zhu}, {Henning}, {Chandler}, {Galv{\'a}n-Madrid},
  {Mac{\'\i}as}, {Anglada}, {Linz}, {Osorio}, {Rodr{\'\i}guez}, {Testi},
  {Torrelles}, {P{\'e}rez}, \& {Liu}}]{Carrasco-Gonzalez_2019}
{Carrasco-Gonz{\'a}lez}, C., {Sierra}, A., {Flock}, M., {et~al.} 2019, \apj,
  883, 71, \dodoi{10.3847/1538-4357/ab3d33}

\bibitem[{{CASA Team} {et~al.}(2022){CASA Team}, {Bean}, {Bhatnagar}, {Castro},
  {Donovan Meyer}, {Emonts}, {Garcia}, {Garwood}, {Golap}, {Gonzalez Villalba},
  {Harris}, {Hayashi}, {Hoskins}, {Hsieh}, {Jagannathan}, {Kawasaki},
  {Keimpema}, {Kettenis}, {Lopez}, {Marvil}, {Masters}, {McNichols},
  {Mehringer}, {Miel}, {Moellenbrock}, {Montesino}, {Nakazato}, {Ott}, {Petry},
  {Pokorny}, {Raba}, {Rau}, {Schiebel}, {Schweighart}, {Sekhar}, {Shimada},
  {Small}, {Steeb}, {Sugimoto}, {Suoranta}, {Tsutsumi}, {van Bemmel},
  {Verkouter}, {Wells}, {Xiong}, {Szomoru}, {Griffith}, {Glendenning}, \&
  {Kern}}]{CASA_2022}
{CASA Team}, {Bean}, B., {Bhatnagar}, S., {et~al.} 2022, \pasp, 134, 114501,
  \dodoi{10.1088/1538-3873/ac9642}

\bibitem[{{Codella} {et~al.}(2014){Codella}, {Cabrit}, {Gueth}, {Podio},
  {Leurini}, {Bachiller}, {Gusdorf}, {Lefloch}, {Nisini}, {Tafalla}, \&
  {Yvart}}]{Codella_2014}
{Codella}, C., {Cabrit}, S., {Gueth}, F., {et~al.} 2014, \aap, 568, L5,
  \dodoi{10.1051/0004-6361/201424103}

\bibitem[{{Doi} \& {Kataoka}(2021)}]{Doi_Kataoka_2021}
{Doi}, K., \& {Kataoka}, A. 2021, \apj, 912, 164,
  \dodoi{10.3847/1538-4357/abe5a6}

\bibitem[{{Drazkowska} {et~al.}(2022){Drazkowska}, {Bitsch}, {Lambrechts},
  {Mulders}, {Harsono}, {Vazan}, {Liu}, {Ormel}, {Kretke}, \&
  {Morbidelli}}]{Drazkowska_2022}
{Drazkowska}, J., {Bitsch}, B., {Lambrechts}, M., {et~al.} 2022, arXiv
  e-prints, arXiv:2203.09759.
\newblock \doarXiv{2203.09759}

\bibitem[{{Dullemond} {et~al.}(2018){Dullemond}, {Birnstiel}, {Huang},
  {Kurtovic}, {Andrews}, {Guzm{\'a}n}, {P{\'e}rez}, {Isella}, {Zhu}, {Benisty},
  {Wilner}, {Bai}, {Carpenter}, {Zhang}, \& {Ricci}}]{Dullemond_2018}
{Dullemond}, C.~P., {Birnstiel}, T., {Huang}, J., {et~al.} 2018, \apjl, 869,
  L46, \dodoi{10.3847/2041-8213/aaf742}

\bibitem[{{Dunham} {et~al.}(2015){Dunham}, {Allen}, {Evans},
  {Broekhoven-Fiene}, {Cieza}, {Di Francesco}, {Gutermuth}, {Harvey},
  {Hatchell}, {Heiderman}, {Huard}, {Johnstone}, {Kirk}, {Matthews}, {Miller},
  {Peterson}, \& {Young}}]{Dunham_2015}
{Dunham}, M.~M., {Allen}, L.~E., {Evans}, Neal~J., I., {et~al.} 2015, \apjs,
  220, 11, \dodoi{10.1088/0067-0049/220/1/11}

\bibitem[{{Flores} {et~al.}(2021){Flores}, {Duch{\^e}ne}, {Wolff}, {Villenave},
  {Stapelfeldt}, {Williams}, {Pinte}, {Padgett}, {Connelley}, {van der Plas},
  {M{\'e}nard}, \& {Perrin}}]{Flores_2021}
{Flores}, C., {Duch{\^e}ne}, G., {Wolff}, S., {et~al.} 2021, \aj, 161, 239,
  \dodoi{10.3847/1538-3881/abeb1e}

\bibitem[{{Galli} {et~al.}(2019){Galli}, {Loinard}, {Bouy}, {Sarro},
  {Ortiz-Le{\'o}n}, {Dzib}, {Olivares}, {Heyer}, {Hernandez},
  {Rom{\'a}n-Z{\'u}{\~n}iga}, {Kounkel}, \& {Covey}}]{Galli_2019}
{Galli}, P.~A.~B., {Loinard}, L., {Bouy}, H., {et~al.} 2019, \aap, 630, A137,
  \dodoi{10.1051/0004-6361/201935928}

\bibitem[{{Gr{\"a}fe} {et~al.}(2013){Gr{\"a}fe}, {Wolf}, {Guilloteau},
  {Dutrey}, {Stapelfeldt}, {Pontoppidan}, \& {Sauter}}]{Grafe_2013}
{Gr{\"a}fe}, C., {Wolf}, S., {Guilloteau}, S., {et~al.} 2013, \aap, 553, A69,
  \dodoi{10.1051/0004-6361/201220720}

\bibitem[{{Gullbring} {et~al.}(1998){Gullbring}, {Hartmann}, {Brice{\~n}o}, \&
  {Calvet}}]{Gullbring_1998}
{Gullbring}, E., {Hartmann}, L., {Brice{\~n}o}, C., \& {Calvet}, N. 1998, \apj,
  492, 323, \dodoi{10.1086/305032}

\bibitem[{{Harris} {et~al.}(2020){Harris}, {Millman}, {van der Walt},
  {Gommers}, {Virtanen}, {Cournapeau}, {Wieser}, {Taylor}, {Berg}, {Smith},
  {Kern}, {Picus}, {Hoyer}, {van Kerkwijk}, {Brett}, {Haldane}, {del R{\'\i}o},
  {Wiebe}, {Peterson}, {G{\'e}rard-Marchant}, {Sheppard}, {Reddy}, {Weckesser},
  {Abbasi}, {Gohlke}, \& {Oliphant}}]{Harris_2020}
{Harris}, C.~R., {Millman}, K.~J., {van der Walt}, S.~J., {et~al.} 2020, \nat,
  585, 357, \dodoi{10.1038/s41586-020-2649-2}

\bibitem[{{Harris} {et~al.}(2018){Harris}, {Cox}, {Looney}, {Li}, {Yang},
  {Fern{\'a}ndez-L{\'o}pez}, {Kwon}, {Sadavoy}, {Segura-Cox}, {Stephens}, \&
  {Tobin}}]{Harris_2018}
{Harris}, R.~J., {Cox}, E.~G., {Looney}, L.~W., {et~al.} 2018, \apj, 861, 91,
  \dodoi{10.3847/1538-4357/aac6ec}

\bibitem[{{Hunter}(2007)}]{Hunter_2007}
{Hunter}, J.~D. 2007, Computing in Science and Engineering, 9, 90,
  \dodoi{10.1109/MCSE.2007.55}

\bibitem[{{Johansen} \& {Klahr}(2005)}]{Johansen_2005}
{Johansen}, A., \& {Klahr}, H. 2005, \apj, 634, 1353, \dodoi{10.1086/497118}

\bibitem[{{Kenyon} \& {Hartmann}(1995)}]{Kenyon_1995}
{Kenyon}, S.~J., \& {Hartmann}, L. 1995, \apjs, 101, 117,
  \dodoi{10.1086/192235}

\bibitem[{{Kristensen} \& {Dunham}(2018)}]{Kristensen_2018}
{Kristensen}, L.~E., \& {Dunham}, M.~M. 2018, \aap, 618, A158,
  \dodoi{10.1051/0004-6361/201731584}

\bibitem[{{Lee}(2020)}]{Lee_2020}
{Lee}, C.-F. 2020, \aapr, 28, 1, \dodoi{10.1007/s00159-020-0123-7}

\bibitem[{{Lee} {et~al.}(2022){Lee}, {Codella}, {Ceccarelli}, \&
  {L{\'o}pez-Sepulcre}}]{Lee_2022}
{Lee}, C.-F., {Codella}, C., {Ceccarelli}, C., \& {L{\'o}pez-Sepulcre}, A.
  2022, \apj, 937, 10, \dodoi{10.3847/1538-4357/ac8c28}

\bibitem[{{Lee} {et~al.}(2014){Lee}, {Hirano}, {Zhang}, {Shang}, {Ho}, \&
  {Krasnopolsky}}]{Lee_2014}
{Lee}, C.-F., {Hirano}, N., {Zhang}, Q., {et~al.} 2014, \apj, 786, 114,
  \dodoi{10.1088/0004-637X/786/2/114}

\bibitem[{{Lee} {et~al.}(2017){Lee}, {Li}, {Ho}, {Hirano}, {Zhang}, \&
  {Shang}}]{Lee_2017}
{Lee}, C.-F., {Li}, Z.-Y., {Ho}, P. T.~P., {et~al.} 2017, Science Advances, 3,
  e1602935, \dodoi{10.1126/sciadv.1602935}

\bibitem[{{Lin} {et~al.}(2021){Lin}, {Lee}, {Li}, {Tobin}, \&
  {Turner}}]{Lin_2021}
{Lin}, Z.-Y.~D., {Lee}, C.-F., {Li}, Z.-Y., {Tobin}, J.~J., \& {Turner}, N.~J.
  2021, \mnras, 501, 1316, \dodoi{10.1093/mnras/staa3685}

\bibitem[{{Liu} {et~al.}(2022){Liu}, {Bertrang}, {Flock}, {Rosotti}, {van
  Dishoeck}, {Boehler}, {Facchini}, {Cui}, {Wolf}, \& {Fang}}]{Liu_2022}
{Liu}, Y., {Bertrang}, G. H.~M., {Flock}, M., {et~al.} 2022, Science China
  Physics, Mechanics, and Astronomy, 65, 129511,
  \dodoi{10.1007/s11433-022-1982-y}

\bibitem[{{Long} {et~al.}(2018){Long}, {Pinilla}, {Herczeg}, {Harsono},
  {Dipierro}, {Pascucci}, {Hendler}, {Tazzari}, {Ragusa}, {Salyk}, {Edwards},
  {Lodato}, {van de Plas}, {Johnstone}, {Liu}, {Boehler}, {Cabrit}, {Manara},
  {Menard}, {Mulders}, {Nisini}, {Fischer}, {Rigliaco}, {Banzatti}, {Avenhaus},
  \& {Gully-Santiago}}]{Long_2018}
{Long}, F., {Pinilla}, P., {Herczeg}, G.~J., {et~al.} 2018, \apj, 869, 17,
  \dodoi{10.3847/1538-4357/aae8e1}

\bibitem[{{Lucas} \& {Roche}(1997)}]{Lucas_Roche_1997}
{Lucas}, P.~W., \& {Roche}, P.~F. 1997, \mnras, 286, 895,
  \dodoi{10.1093/mnras/286.4.895}

\bibitem[{{Maret} {et~al.}(2020){Maret}, {Maury}, {Belloche}, {Gaudel},
  {Andr{\'e}}, {Cabrit}, {Codella}, {Lef{\'e}vre}, {Podio}, {Anderl}, {Gueth},
  \& {Hennebelle}}]{Maret_2020}
{Maret}, S., {Maury}, A.~J., {Belloche}, A., {et~al.} 2020, \aap, 635, A15,
  \dodoi{10.1051/0004-6361/201936798}

\bibitem[{{Melis} {et~al.}(2011){Melis}, {Duch{\^e}ne}, {Chomiuk}, {Palmer},
  {Perrin}, {Maddison}, {M{\'e}nard}, {Stapelfeldt}, {Pinte}, \&
  {Duvert}}]{Melis_2011}
{Melis}, C., {Duch{\^e}ne}, G., {Chomiuk}, L., {et~al.} 2011, \apjl, 739, L7,
  \dodoi{10.1088/2041-8205/739/1/L7}

\bibitem[{{Mendigut{\'\i}a} {et~al.}(2013){Mendigut{\'\i}a}, {Brittain},
  {Eiroa}, {Meeus}, {Montesinos}, {Mora}, {Muzerolle}, {Oudmaijer}, \&
  {Rigliaco}}]{Mendigutia_2013}
{Mendigut{\'\i}a}, I., {Brittain}, S., {Eiroa}, C., {et~al.} 2013, \apj, 776,
  44, \dodoi{10.1088/0004-637X/776/1/44}

\bibitem[{{Michel} {et~al.}(2022){Michel}, {Sadavoy}, {Sheehan}, {Looney}, \&
  {Cox}}]{Michel_2022}
{Michel}, A., {Sadavoy}, S.~I., {Sheehan}, P.~D., {Looney}, L.~W., \& {Cox},
  E.~G. 2022, \apj, 937, 104, \dodoi{10.3847/1538-4357/ac905c}

\bibitem[{{Murillo} {et~al.}(2018){Murillo}, {Harsono}, {McClure}, {Lai}, \&
  {Hogerheijde}}]{Murillo_2018}
{Murillo}, N.~M., {Harsono}, D., {McClure}, M., {Lai}, S.~P., \& {Hogerheijde},
  M.~R. 2018, \aap, 615, L14, \dodoi{10.1051/0004-6361/201833420}

\bibitem[{{Muro-Arena} {et~al.}(2018){Muro-Arena}, {Dominik}, {Waters}, {Min},
  {Klarmann}, {Ginski}, {Isella}, {Benisty}, {Pohl}, {Garufi}, {Hagelberg},
  {Langlois}, {Menard}, {Pinte}, {Sezestre}, {van der Plas}, {Villenave},
  {Delboulb{\'e}}, {Magnard}, {M{\"o}ller-Nilsson}, {Pragt}, {Rabou}, \&
  {Roelfsema}}]{Muro-Arena_2018}
{Muro-Arena}, G.~A., {Dominik}, C., {Waters}, L.~B.~F.~M., {et~al.} 2018, \aap,
  614, A24, \dodoi{10.1051/0004-6361/201732299}

\bibitem[{{Nakatani} {et~al.}(2020){Nakatani}, {Liu}, {Ohashi}, {Zhang},
  {Hanawa}, {Chandler}, {Oya}, \& {Sakai}}]{Natakani_2020}
{Nakatani}, R., {Liu}, H.~B., {Ohashi}, S., {et~al.} 2020, \apjl, 895, L2,
  \dodoi{10.3847/2041-8213/ab8eaa}

\bibitem[{{Ohashi} {et~al.}(2022){Ohashi}, {Nakatani}, {Liu}, {Kobayashi},
  {Zhang}, {Hanawa}, \& {Sakai}}]{Ohashi_2022}
{Ohashi}, S., {Nakatani}, R., {Liu}, H.~B., {et~al.} 2022, \apj, 934, 163,
  \dodoi{10.3847/1538-4357/ac794e}

\bibitem[{{Padgett} {et~al.}(1999){Padgett}, {Brandner}, {Stapelfeldt},
  {Strom}, {Terebey}, \& {Koerner}}]{Padgett_1999}
{Padgett}, D.~L., {Brandner}, W., {Stapelfeldt}, K.~R., {et~al.} 1999, \aj,
  117, 1490, \dodoi{10.1086/300781}

\bibitem[{{Pinte} {et~al.}(2016){Pinte}, {Dent}, {M{\'e}nard}, {Hales}, {Hill},
  {Cortes}, \& {de Gregorio-Monsalvo}}]{Pinte_2016}
{Pinte}, C., {Dent}, W.~R.~F., {M{\'e}nard}, F., {et~al.} 2016, \apj, 816, 25,
  \dodoi{10.3847/0004-637X/816/1/25}

\bibitem[{{Pinte} {et~al.}(2009){Pinte}, {Harries}, {Min}, {Watson},
  {Dullemond}, {Woitke}, {M{\'e}nard}, \& {Dur{\'a}n-Rojas}}]{Pinte_2009}
{Pinte}, C., {Harries}, T.~J., {Min}, M., {et~al.} 2009, \aap, 498, 967,
  \dodoi{10.1051/0004-6361/200811555}

\bibitem[{{Pinte} {et~al.}(2006){Pinte}, {M{\'e}nard}, {Duch{\^e}ne}, \&
  {Bastien}}]{Pinte_2006}
{Pinte}, C., {M{\'e}nard}, F., {Duch{\^e}ne}, G., \& {Bastien}, P. 2006, \aap,
  459, 797, \dodoi{10.1051/0004-6361:20053275}

\bibitem[{{Podio} {et~al.}(2020){Podio}, {Garufi}, {Codella}, {Fedele},
  {Bianchi}, {Bacciotti}, {Ceccarelli}, {Favre}, {Mercimek}, {Rygl}, \&
  {Testi}}]{Podio_2020}
{Podio}, L., {Garufi}, A., {Codella}, C., {et~al.} 2020, \aap, 642, L7,
  \dodoi{10.1051/0004-6361/202038952}

\bibitem[{{Robitaille} {et~al.}(2007){Robitaille}, {Whitney}, {Indebetouw}, \&
  {Wood}}]{Robitaille_2007}
{Robitaille}, T.~P., {Whitney}, B.~A., {Indebetouw}, R., \& {Wood}, K. 2007,
  \apjs, 169, 328, \dodoi{10.1086/512039}

\bibitem[{{Rodmann} {et~al.}(2006){Rodmann}, {Henning}, {Chandler}, {Mundy}, \&
  {Wilner}}]{Rodmann_2006}
{Rodmann}, J., {Henning}, T., {Chandler}, C.~J., {Mundy}, L.~G., \& {Wilner},
  D.~J. 2006, \aap, 446, 211, \dodoi{10.1051/0004-6361:20054038}

\bibitem[{{Rosotti} {et~al.}(2020){Rosotti}, {Teague}, {Dullemond}, {Booth}, \&
  {Clarke}}]{Rosotti_2020}
{Rosotti}, G.~P., {Teague}, R., {Dullemond}, C., {Booth}, R.~A., \& {Clarke},
  C.~J. 2020, \mnras, 495, 173, \dodoi{10.1093/mnras/staa1170}

\bibitem[{{Rota} {et~al.}(2022){Rota}, {Manara}, {Miotello}, {Lodato},
  {Facchini}, {Koutoulaki}, {Herczeg}, {Long}, {Tazzari}, {Cabrit}, {Harsono},
  {M{\'e}nard}, {Pinilla}, {van der Plas}, {Ragusa}, \& {Yen}}]{Rota_2022}
{Rota}, A.~A., {Manara}, C.~F., {Miotello}, A., {et~al.} 2022, \aap, 662, A121,
  \dodoi{10.1051/0004-6361/202141035}

\bibitem[{{Sakai} {et~al.}(2017){Sakai}, {Oya}, {Higuchi}, {Aikawa}, {Hanawa},
  {Ceccarelli}, {Lefloch}, {L{\'o}pez-Sepulcre}, {Watanabe}, {Sakai}, {Hirota},
  {Caux}, {Vastel}, {Kahane}, \& {Yamamoto}}]{Sakai_2017}
{Sakai}, N., {Oya}, Y., {Higuchi}, A.~E., {et~al.} 2017, \mnras, 467, L76,
  \dodoi{10.1093/mnrasl/slx002}

\bibitem[{{Sheehan} {et~al.}(2022){Sheehan}, {Tobin}, {Li}, {van't Hoff},
  {J{\o}rgensen}, {Kwon}, {Looney}, {Ohashi}, {Takakuwa}, {Williams}, {Aso},
  {Gavino}, {Gregorio-Monsalvo}, {Han}, {Lee}, {Plunkett}, {Sharma}, {Aikawa},
  {Lai}, {Lee}, {Lin}, {Saigo}, {Tomida}, \& {Yen}}]{Sheehan_2022}
{Sheehan}, P.~D., {Tobin}, J.~J., {Li}, Z.-Y., {et~al.} 2022, \apj, 934, 95,
  \dodoi{10.3847/1538-4357/ac7a3b}

\bibitem[{{Silsbee} {et~al.}(2022){Silsbee}, {Akimkin}, {Ivlev}, {Testi},
  {Gong}, \& {Caselli}}]{Silsbee_2022}
{Silsbee}, K., {Akimkin}, V., {Ivlev}, A.~V., {et~al.} 2022, \apj, 940, 188,
  \dodoi{10.3847/1538-4357/ac978b}

\bibitem[{{Teague} {et~al.}(2019){Teague}, {Bae}, \& {Bergin}}]{Teague_2019}
{Teague}, R., {Bae}, J., \& {Bergin}, E.~A. 2019, \nat, 574, 378,
  \dodoi{10.1038/s41586-019-1642-0}

\bibitem[{{Tobin} {et~al.}(2008){Tobin}, {Hartmann}, {Calvet}, \&
  {D'Alessio}}]{Tobin_2008}
{Tobin}, J.~J., {Hartmann}, L., {Calvet}, N., \& {D'Alessio}, P. 2008, \apj,
  679, 1364, \dodoi{10.1086/587683}

\bibitem[{{van't Hoff} {et~al.}(2020){van't Hoff}, {Harsono}, {Tobin},
  {Bosman}, {van Dishoeck}, {J{\o}rgensen}, {Miotello}, {Murillo}, \&
  {Walsh}}]{vant_Hoff_2020}
{van't Hoff}, M. L.~R., {Harsono}, D., {Tobin}, J.~J., {et~al.} 2020, \apj,
  901, 166, \dodoi{10.3847/1538-4357/abb1a2}

\bibitem[{{Villenave} {et~al.}(2020){Villenave}, {M{\'e}nard}, {Dent},
  {Duch{\^e}ne}, {Stapelfeldt}, {Benisty}, {Boehler}, {van der Plas}, {Pinte},
  {Telkamp}, {Wolff}, {Flores}, {Lesur}, {Louvet}, {Riols}, {Dougados},
  {Williams}, \& {Padgett}}]{Villenave_2020}
{Villenave}, M., {M{\'e}nard}, F., {Dent}, W.~R.~F., {et~al.} 2020, \aap, 642,
  A164, \dodoi{10.1051/0004-6361/202038087}

\bibitem[{{Villenave} {et~al.}(2022){Villenave}, {Stapelfeldt}, {Duch{\^e}ne},
  {M{\'e}nard}, {Lambrechts}, {Sierra}, {Flores}, {Dent}, {Wolff}, {Ribas},
  {Benisty}, {Cuello}, \& {Pinte}}]{Villenave_2022}
{Villenave}, M., {Stapelfeldt}, K.~R., {Duch{\^e}ne}, G., {et~al.} 2022, \apj,
  930, 11, \dodoi{10.3847/1538-4357/ac5fae}

\bibitem[{{Weidenschilling}(1977)}]{Weidenschilling_1977}
{Weidenschilling}, S.~J. 1977, \mnras, 180, 57, \dodoi{10.1093/mnras/180.2.57}

\bibitem[{{Wolf} {et~al.}(2003){Wolf}, {Padgett}, \& {Stapelfeldt}}]{Wolf_2003}
{Wolf}, S., {Padgett}, D.~L., \& {Stapelfeldt}, K.~R. 2003, \apj, 588, 373,
  \dodoi{10.1086/374041}

\bibitem[{{Wolf} {et~al.}(2008){Wolf}, {Schegerer}, {Beuther}, {Padgett}, \&
  {Stapelfeldt}}]{Wolf_2008}
{Wolf}, S., {Schegerer}, A., {Beuther}, H., {Padgett}, D.~L., \& {Stapelfeldt},
  K.~R. 2008, \apjl, 674, L101, \dodoi{10.1086/529188}

\bibitem[{{Wolff} {et~al.}(2021){Wolff}, {Duch{\^e}ne}, {Stapelfeldt},
  {M{\'e}nard}, {Flores}, {Padgett}, {Pinte}, {Villenave}, {van der Plas}, \&
  {Perrin}}]{Wolff_2021}
{Wolff}, S.~G., {Duch{\^e}ne}, G., {Stapelfeldt}, K.~R., {et~al.} 2021, \aj,
  161, 238, \dodoi{10.3847/1538-3881/abeb1d}

\bibitem[{{Youdin} \& {Lithwick}(2007)}]{Youdin_2007}
{Youdin}, A.~N., \& {Lithwick}, Y. 2007, \icarus, 192, 588,
  \dodoi{10.1016/j.icarus.2007.07.012}

\bibitem[{{Zhang} {et~al.}(2021){Zhang}, {Launhardt}, {Liu}, {Tobin}, \&
  {Henning}}]{Zhang_2020}
{Zhang}, C.-P., {Launhardt}, R., {Liu}, Y., {Tobin}, J.~J., \& {Henning}, T.
  2021, \aap, 646, A18, \dodoi{10.1051/0004-6361/202039536}

\end{thebibliography}
\bibliographystyle{aasjournal}

\end{document}